\documentclass[11pt]{article}

\usepackage{amssymb,latexsym, amsmath}
\usepackage[colorlinks]{hyperref}
\usepackage{color}

\makeatletter \@addtoreset{figure}{section}
\def\thefigure{\thesection.\@arabic\c@figure} \def\fps@figure{h, t}
\@addtoreset{table}{bsection}
\def\thetable{\thesection.\@arabic\c@table} \def\fps@table{h, t}
\@addtoreset{equation}{section}

\makeatother

\topmargin by -\headsep \textheight 8.9in \oddsidemargin 0pt
\evensidemargin \oddsidemargin \marginparwidth 0.5in \textwidth
6.5 truein

\newtheorem{thm}{Theorem}[section]

\begin{document}

\title{The Complex Geometry of Weak Piecewise Smooth Solutions of Integrable
Nonlinear PDE's of Shallow Water and Dym Type\footnote{PACS
numbers 05.45.Yv, 03.40.Gc, 11.10.Ef, 68.10.-m, AMS Subject
Classification 58F07, 70H99, 76B15}}
\author {Mark S. Alber\thanks{Research partially supported by NSF grant DMS
9626672 and NATO grant CRG 950897.} \\Department of Mathematics,
Stanford University, Building 380, MC 2125, Stanford, CA 94305 \\
{\footnotesize malber@math.stanford.edu}\\ and\\ Department of
Mathematics, University of Notre Dame, Notre Dame, IN 46556 \\
{\footnotesize Mark.S.Alber.1@nd.edu} \and Roberto
Camassa\thanks{Research supported in part by US DOE CCPP and BES
programs and NATO grant CRG 950897}\\ Department of Mathematics,
University of North Carolina\\ Chapel Hill, NC 27599 \\ and \\
Center for Nonlinear Studies and Theoretical Division \\ Los
Alamos National Laboratory, Los Alamos, NM 87545 \\ {\footnotesize
camassa@math.unc.edu} \and Yuri N. Fedorov\thanks{Research
supported by INTAS grant 97-10771 and, in part, by the Center for
Applied Mathematics, University of Notre Dame}
\\Department of Mathematics and Mechanics, Moscow Lomonosov University, \\\
Moscow 119 899, Russia
\\ {\footnotesize fedorov@mech.math.msu.su}
\and Darryl D. Holm\thanks{Research supported in part by US DOE
CCPP and BES programs}\\ Center for Nonlinear Studies and
Theoretical Division \\ Los Alamos National Laboratory, Los
Alamos, NM 87545\\ {\footnotesize dholm@lanl.gov} \and Jerrold E.
Marsden\thanks{Research partially supported by the California
Institute of Technology and NSF grant DMS 9802106}\\ Control and
Dynamical Systems 107-81, California Institute of Technology \\
Pasadena, CA 91125\\ {\footnotesize  marsden@cds.caltech.edu} \\}
\date{April 30, 2001}

\maketitle

\newpage
{\it Short title}: The Complex Geometry of Piecewise Solutions
%%%%%%%%%%%%%%%%%%%%%%%%%%%%%%%%%%%%%%%%%%%%%%%%%%%%%%%%%%%%%%

\begin{abstract}
An extension of the algebraic-geometric method for nonlinear
integrable PDE's is shown to lead to new piecewise smooth weak
solutions of a class of $N$-component systems of nonlinear
evolution equations. This class includes, among others, equations from the
Dym and shallow water equation hierarchies. The main goal of the
paper is to give explicit theta-functional solutions of these nonlinear PDE's,
which are associated to nonlinear subvarieties of hyperelliptic Jacobians.

The main results of the present paper are twofold. First, we exhibit
some of the special features of integrable PDE's that admit piecewise
smooth weak solutions, which make them different from equations whose
solutions are globally meromorphic, such as the KdV equation. Second,  we
blend the techniques of algebraic geometry and weak solutions of PDE's to
gain further insight into, and explicit formulas for, piecewise-smooth
finite-gap solutions.

The basic technique used to achieve these aims is rather
different from earlier papers dealing with peaked solutions.
First, profiles of the finite-gap piecewise smooth solutions are
linked to certain finite dimensional billiard dynamical systems
and ellipsoidal billiards. Second, after reducing the solution of
certain finite dimensional Hamiltonian systems on Riemann surfaces to
the solution of a nonstandard Jacobi inversion problem, this is
resolved by introducing new parametrizations.

Amongst other natural consequences of the algebraic-geometric
approach, we find finite dimensional integrable Hamiltonian
dynamical systems describing the motion of peaks in the finite-gap
as well as the limiting (soliton) cases, and solve them exactly. The
dynamics of the peaks is also obtained by using Jacobi inversion
problems. Finally, we relate our method to the shock wave approach
for weak solutions of wave equations  by determining jump
conditions at the peak location.
\end{abstract}

\tableofcontents
%%%%%%%%%%%%%%%%%%%%%%%%%%%%%%%%%%%%%%%%%%%%%%%%%%%%%%%%%%%%%%

\section{Introduction}

An important feature of many integrable nonlinear evolution
equations is the nature of their soliton solutions. There are many
examples of such solutions found in a variety of physical
applications, such as nonlinear optics and water wave equations.
Nonsmooth soliton solutions of integrable equations are now well
known, and include solutions of the shallow water equation
(\ref{SW-sw-eqn}) with peaks, the points at which their spatial
derivative changes sign (see Camassa and Holm [1993] and Camassa,
Holm and Hyman [1994]). It was noted in Alber {\it et al.} [1994,
1995, 1999] that the spatial structure of these ``peakon" and
finite-gap piecewise smooth weak solutions are closely related to
finite dimensional integrable billiard systems.

The purpose of this paper is twofold. First, we exhibit some of the
special features of integrable PDE's that admit piecewise smooth
weak solutions, which make them different from equations whose
solutions are globally meromorphic, such as the KdV equation.
Second,  we blend the techniques of algebraic geometry and weak
solutions of PDE's to gain further insight into finite-gap
solutions that are piecewise smooth.
%

%%%%%%%%%%%%%%%%%%%%%%%%%%%%%%%%%%%%%%%%%%%%%%%%%%%%%%%%%%%%%%

\paragraph{Some history.}

Camassa and Holm [1993] described classes of $n$-peakon solutions
for an integrable equation in the context of a model for shallow
water theory. This work (see also Camassa, Holm and Hyman [1994])
contains many other facts about these equations as well, such as a
Hamiltonian derivation of the equation,
the associated linear isospectral eigenvalue problem and its
discrete spectrum corresponding to the peakons,
a steepening lemma important for understanding how solutions lose
regularity, numerical stability, etc. Of particular interest to us
is their description of the dynamics of the peakons in terms of a
finite-dimensional completely integrable Hamiltonian system. In
other words, each peakon solution can be associated with a
mechanical system of moving particles. Calogero [1995] and
Calogero and Francoise [1996] further extended the class of
mechanical systems of this type.

It is well-known (see, for example, Ablowitz and Segur [1981]),
that solitons and quasi-periodic solutions of most classical
integrable equations can be obtained by using the inverse
scattering transform (IST) method. This is done by establishing a
connection with an isospectral eigenvalue problem for an
associated operator that is often a Schr\"{o}dinger operator. In
some cases it involves a potential in the form of an entire
function of the spectral parameter. Such an operator is called an
{\it energy-dependent} Schr\"{o}dinger operator. The scattering
problem for the operators of this type was studied by Jaulent
[1972] and Jaulent and Jean [1976].

On the other hand, in connection with certain $N$-component
systems of integrable evolution equations, Antonowicz and Fordy
[1989]  investigated certain energy dependent {\bf scalar}
Schr\"{o}dinger operators. Using this formalism, they obtained
multi-Hamiltonian structures for this class of systems.

Later, Alber {\it et al.} [1994, 1995, 1999] showed that in case
of certain potentials, a limiting procedure can be applied to
generic solutions, which results in solutions with peaks. The
latter were related to finite dimensional integrable dynamical
systems with reflections and were termed {\it piecewise-smooth
solutions}, a terminology that hereafter we will adopt. This
relation provides an efficient route to the study of finite-gap
and piecewise soliton solutions of nonlinear PDE's. The approach
is based on studying finite dimensional Hamiltonian systems on
certain Riemann surfaces and can be used for a number of equations
including the shallow water equation, the Dym type equation, as
well as certain $N$-component systems  and equations in their
hierarchies.

Finite-gap solutions of the Dym equation were studied in Dmitrieva
[1993a] and Novikov [1999] by making use of a connection with the
KdV equation and with the aid of additional phase functions.
Soliton solutions of Dym type equations were studied in Dmitrieva
[1993b]. Periodic solutions of the shallow water equation were
discussed in McKean and Constantin [1999]. The papers by Beals
{\it et al.} [1998, 1999, 2000] used Stieltjes' theorem on
continued fractions and the classical moment problem for  studying
multi-peakon solutions of the (\ref{SW-sw-eqn}) equation.
Multi-peakon solutions have also been derived in Camassa [2000] by
Gram--Schmidt orthogonalization.
%%%%%%%%%%%%%%%%%%%%%%%%%%%%%%%%%%%%%%%%%%%%%%%%%%%%%%%%%%%%%%

\paragraph{The main results of this paper.}
While our techniques are rather general and can be applied to
large classes of $N$-component systems, we shall illustrate them
in detail for two specific integrable PDE's. One of these
equations is a member of the Dym hierarchy that has been studied
by, amongst others, Kruskal [1975], Cewen [1990], Hunter and Zheng
[1994] and Alber {\it et al.} [1995, 1999]. Using subscript
notation for partial derivatives, this equation is
%-----------------
\begin{equation}
\tag{HD} \label{dym-sw-eqn} U_{xxt}+2U_x U_{xx}+U U_{xxx}-2\kappa
U_x=0 \, .
\end{equation}
The other equation, derived from the Euler equations of
hydrodynamics in a shallow water framework by Camassa and Holm [1993], is
\begin{equation}
\tag{SW} \label{SW-sw-eqn} U_t+3UU_x = U_{xxt}+2U_x U_{xx}+U
U_{xxx}-2\kappa U_x\,.
\end{equation}
In both equations, the dependent variable $U(x,t)$ may be
interpreted as a horizontal fluid velocity and $\kappa$ is a
parameter.

Under appropriate boundary conditions, applying the limit
$\kappa\rightarrow 0 $ to (\ref{SW-sw-eqn}) leads to an equation
that has peaked solutions. For equation (\ref{dym-sw-eqn}), such
solutions exist also for $\kappa \neq 0 $ (for example periodic
and finite-gap peaked solutions).

By using the method of generating equations for nonlinear
integrable PDE's, we reduce the equations to a Jacobi inversion
problem associated with hyperelliptic curves. The solutions
$U(x,t)$ themselves are given by trace formulae,  i.e., sums of
coordinates of points on such curves.

An important feature is that the corresponding Abel--Jacobi
mapping is not a standard one. First of all, the holomorphic
differentials that are involved do not form a complete set of such
differentials on a hyperelliptic curve. Second, it involves a
meromorphic differential. As a result, the  image of the mapping
turns out to be a  non-Abelian subvariety (a stratum) of a {\it
generalized} Jacobian. This also implies that the $x$- and $t$-flows
of (\ref{dym-sw-eqn}) and (\ref{SW-sw-eqn}) are essentially
nonlinear,  i.e., they are not translationally invariant. Seen
from the viewpoint of algebraic geometry, these nonstandard
aspects constitute the main difference between shallow water and
Dym type equations, and equations of KdV type and more generally
equations from the whole KP hierarchy which lead to standard
Abel--Jacobi mappings.

   The basic technique of the present paper is rather
different from earlier papers dealing with peaked solutions.
First, profiles of the finite-gap piecewise-smooth solutions are
linked to certain finite dimensional billiard dynamical systems
and ellipsoidal billiards. Second, after reducing the solution of
the finite dimensional Hamiltonian systems on Riemann surfaces to
the solution of a nonstandard Jacobi inversion problem, it is
resolved by introducing new parametrizations.

The philosophy that ``justifies'' procedures of this sort is that,
in the end, by using the trace formulae, {\it we obtain weak
solutions of the PDE's (\ref{dym-sw-eqn}) and (\ref{SW-sw-eqn}) in
the spacetime sense}. This is regarded as equivalent to the
validity of Hamilton's principle for these PDE's and is taken as a
fundamental criterion for the definition of their solutions. It is
worth emphasizing that Hamilton's principle naturally leads to
weak solutions in the {\it spacetime} sense (and not in the
spatial sense alone). We might also remark that even for
billiards, one has to be careful about the sense in which
solutions are interpreted. In the case of a point particle
bouncing off a wall, for example, the equations of motion
themselves do not rigorously make sense at the collision; what
does make sense is the fundamental principle of Hamilton. This
point of view of course is not new --- see, e.g., Young [1969] and
Kane et al [1999].

\paragraph{The contents of the paper.}
In \S2, basic trace formulae and $\mu$-variable representations
are used to establish a connection between solutions of the
nonlinear equations  and finite dimensional Hamiltonian systems on
Riemann surfaces. These representations describe finite-gap and
soliton type solutions, as well as mixed soliton--finite-gap
solutions. Then, solving the Hamiltonian systems is reduced to
Jacobi inversion problems with meromorphic differentials. These
inversion problems are solved by introducing a new
parameterization that yields a Hamiltonian flow on a nonlinear
subvariety of the Jacobi variety. The approach of recurrence
chains used in this section is demonstrated in detail in the case
of Dym-type equations.

In \S3 the geodesic motion and motion in the field of a Hooke
potential on an ellipsoid $\tilde Q$ are linked, at any fixed time
$t$, to finite-gap solutions of  (\ref{dym-sw-eqn}) and
(\ref{SW-sw-eqn}) equations respectively through trace formulae.
In \S4 it is  shown how peaked finite-gap solutions of
(\ref{dym-sw-eqn}) and (\ref{SW-sw-eqn}) equations arise in the
particular limit of smooth solutions. Based on this, a connection
to ellipsoidal and hyperbolic billiards is used to construct the
{\it peak} solutions of equations (\ref{dym-sw-eqn}) and
(\ref{SW-sw-eqn}) in the form of an infinite sequence of pieces,
corresponding to the segments between impacts, glued together
along peaks. The motion between impacts in the  billiard problems
is made linear on generalized Jacobians of hyperelliptic curves.

By solving the corresponding generalized Jacobi inversion problem,
we find theta-function solutions to the billiards, which thereby
enables us to describe explicit peak solutions for the above PDE.
We then extend the analysis from fixed-time peak solutions to
time-dependent ones and show that the latter are described by an
infinite number of meromorphic pieces in $x$ and $t$ that are
glued along peak lines (surfaces) where the solution has
discontinuous derivatives in the dependent variables. We give
theta-function expressions for the pieces and the peak surfaces.
These formulae may be useful for stability analysis as well as for
numerical investigations of the perturbed nonlinear PDE's.

In \S5  the Hamiltonian structure for the motion of the peaks of
the finite-gap piecewise-solutions is obtained by using
algebraic-geometric methods. Lastly, in \S6  we relate our method
to the shock wave approach for weak solutions of wave equations by
determining jump conditions at the shock location.

%%%%%%%%%%%%%%%%%%%%%%%%%%%%%%%%%%%%%%%%%%%%%%%%%%%%%%%%%%%%%%

\section{Finite-gap solutions}
\setcounter{equation}{0} In this section we will show that even on
the level of finite-gap solutions, there are crucial differences
between the KdV equation case and equations (\ref{dym-sw-eqn}) or
(\ref{SW-sw-eqn}). The same method can be applied to other
equations forming the HD and SW hierarchy as well as to
$N$-component systems of nonlinear evolution equations which have
associated with them energy dependent Schr\"{o}dinger operators
(see Alber {\it et al.} [1997]) .

We will start by describing the algebraic geometrical structure of
finite-gap solutions of equations (\ref{dym-sw-eqn}) and
(\ref{SW-sw-eqn}) related to a hyperelliptic curve of genus $n$,
also called $n$-gap solutions. The same method can be applied also
to the other equations forming the HD and SW hierarchy.

For the HD equation such solutions were obtained in terms of
theta-functions by Dmitrieva [1993a] (see also Dmitrieva [1993b])
and Novikov [1999]. For equation (\ref{SW-sw-eqn}) on a circle,
the problem was discussed in Constantin and McKean [1999].

\paragraph{Lax pairs and recurrence chains.} We now use the recurrence
chain approach to develop a basic trace formula which establishes
a connection between solutions of equation (\ref{dym-sw-eqn}) and
finite dimensional Hamiltonian systems on Riemann surfaces,
written in the so-called $\mu$-variables representation. This
representation describes finite-gap solutions, as well as their
limiting forms of soliton-type.  This representation also yields
the existence of peakons in a special limiting case. For
definiteness, we concentrate here on equation (\ref{dym-sw-eqn}).
Analogous results are available in the case of equation
(\ref{SW-sw-eqn}) (for details see Alber {\it et al.} [1994,
1995]).

The hierarchy of Dym equations is obtained from the Lax equations
$$ \frac{\partial}{\partial t_n}L=[L,A_n], \qquad n\in{\mathbb N},
\quad L= -\frac{\partial^2}{\partial x^2} + V(E,x,t_n), $$ where
the potential $V(E,x,t_n)$ is written in terms of a  complex
parameter $E$ in the form
\begin{equation}
\label{v} V(x,t_n,E)=\frac{M(x,t_n)}{2E}\, ,
\end{equation}
for a function $M(x,t_n)$ to be determined below. Assuming $[L,
A_n]$ to be a scalar operator, we choose $A_n=B_n\partial_x-\frac
12 B_{n}'$ for some function $B_n(E,x,t_n)$ and obtain the
following sequence of equations for $V$,
\begin{equation}
\label{gneq} \frac{\partial V}{\partial t_n}
=-\frac{1}{2}\frac{\partial^3 B_n}{\partial x^3} +2\frac{\partial
B_n}{\partial x} V +B_n\frac{\partial V}{\partial x} \,.
\end{equation}
Now we choose $B_n$ to be a polynomial in $E$ of degree $n$:
\begin{equation} \label{be}
B_n(x,t,E) = b_0\, \prod_{k=1}^n(E-\mu_k(x,t))=\sum_{k=0}^n
b_{n-k}(x,t) \, E^k. \,
\end{equation}
Substituting the expressions (\ref{v}) and (\ref{be}) into the
generating equation (\ref{gneq}) and equating like powers of $E$,
we obtain a recurrence chain for coefficients of $B(x,t)$ which
yields $n$-th equation of the Dym hierarchy.
For example, putting $t_1=t$ and choosing $n=1$, $B_1(x,t,E) =
b_0(x,t)E + b_1(x,t)$ yields the following chain
\begin{eqnarray} \label{ch0} E^{1} &:& -b_0^{'''}  = 0
\nonumber\\ E^{0} &:& - b_1^{'''} + 2b_0' M + b_0 M' = 0
\nonumber\\ E^{-1} &:& 2b_1' M + b_1 M' = \frac{\partial
M}{\partial t} .
\end{eqnarray}
After setting $b_0 = 1$ and using (\ref{v}), we get
\begin{align}
        M' &= b_1^{'''} \nonumber\\
       2b_1' M + b_1 M' & = \frac{\partial M}{\partial t}. \label{d1}
\end{align}
The first equation defines $b_1$ in terms of $M$, $M  = b_1 ^{''}+
\kappa $, with $\kappa$ a constant. Renaming $b_1 = - U$, so that
\begin{equation}
\label{MU} M=-U''+\kappa,
\end{equation}
and putting this into the second equation of the set (\ref{d1})
results in equation (\ref{dym-sw-eqn}). (For further details about
the hierarchies of (\ref{dym-sw-eqn}) and (\ref{SW-sw-eqn}), see
for example Alber {\it et al.} [1994, 1995, 1999].) The method of
generating equations is due to S. Alber [1979] and another
exposition of it can be found in Alber {\it et al.} [1985, 1997].

We call (\ref{gneq}) the ``dynamical generating equations,"
because it generates a hierarchy of equations governing the
dynamics of the dependent variable $M(x,t)$.
\medskip

\noindent{\bf Remark.} The flows where $B_n$ is a polynomial $E$, as
in the definition~(\ref{be}) and in the example above, will in general lead
to nonlocal equations, i.e., the evolution equation
for $M$ involves terms that depend on nonlocal operators
acting on combinations of $M$ and its derivatives.
This can be seen, for instance, in equation~(\ref{d1})
where both $b_1$ and $b'_1$ require inverting~(\ref{MU}) to
write $U$ in terms of $M$. Thus, flows generated
by polynomials $B_n$ in $E$ should properly classified
as integro-differential evolution equations, rather than PDE's. In
contrast, the choice of  polynomials in $1/E$ for $B_n$ leads to flows
that are local, i.e., $M_t$ only depends
on combinations $M$ and its (spatial) derivatives, and these
flows are proper PDE's.
This feature of equations of Dym (\ref{dym-sw-eqn}) or
shallow water (\ref{SW-sw-eqn}) type is
somewhat different from other completely integrable PDE's
like the KdV or Sine-Gordon equation. Equations (\ref{dym-sw-eqn}) and
(\ref{SW-sw-eqn}) possess ``open ended''
hierarchies:  the recurrence chain can be extended from
negative to positive powers of $E$,  by choosing  $B_n$ in
(\ref{gneq}) to be a rational function of the parameter $E$. The case
when the chain includes only negative powers of $E$ is in fact the one
most studied in the literature (see, e.g., Dimitrieva [1993a], Novikov
[1999] for the case of Dym equation).
\medskip

Now let us consider the stationary flow for the $n$-th equation of
the hierarchy, which is obtained by dropping the time derivative
of $V$ in the left-hand side of (\ref{gneq}). By definition
   stationary equation describes a finite-dimensional system for
the coefficients of $B_n$ and is equivalent to the $2\times 2$ Lax
pair
\begin{gather}
\frac{\partial}{\partial x} W_n(E)=-[W_n(E),{\cal L}(E)] , \quad
\mbox{or} \quad \left[ \frac{\partial}{\partial x}+{\cal
L}(E),W_n(E)\right]=0, \label{Lax-stationary} \\
W_n(E)=\begin{pmatrix}  -\frac 12 B_n' & B_n \\
                       -\frac12 B_{n}''+ B_n \frac M E &  \frac 12 B_{n}'
\end{pmatrix},  \quad
{\cal L}=\begin{pmatrix}  0 & 1 \\
                       \frac M E  &  0 \end{pmatrix} . \nonumber
\end{gather}
The matrix $W_n(E)$ undergoes an isospectral deformation. Hence
the spectral curve $$\Gamma=\{|W_n(E)-z I|=0\}$$ is an invariant of
the stationary flow. The curve is hyperelliptic and can be
represented in the form
\begin{equation} \label{muw1}
\Gamma=\{w^2=\mu C(\mu)\}
\end{equation}
where $z=w E$ and
\begin{equation}
\label{sgneq} C(E)= E\left(-B_n^{''}B_n + \frac{1}{2} B_n^{\prime
2}\right)+B_n^2 M .
\end{equation}
Since $B_n$ is a polynomial of degree $n$, $C(E)$ becomes
a polynomial of degree (at most) $2n$:
\begin{equation}
   C(E)=\sum_{j=0}^{2n} C_j \, E^j =C_{2n}\, \prod_{k=0}^{2n}(E-m_k),
\label{sc}
\end{equation}
for some constants $m_k$, $k=1,\dots,2n$. In this case the curve
$\Gamma$ has genus $n$ and we set coefficient $C_{2n}$ to be a
negative number: $C_{2n} \equiv -L_0^2$.
We shall refer to (\ref{sgneq}) as the {\it stationary generating
equation}.

   Equating like-powers of $E$ in both sides of the stationary generating
equation yields first integrals
\begin{equation}\label{ch}
\begin{aligned} E^{2n} &: C_{2n}&=& -b_1^{''}+ M , \\
E^{2n-1} &:  C_{2n-1} &=&- b_1 b_1^{''}- b_2^{''}+\frac{1}{2}
(b_1')^2 + 2 b_1 M,\\ E^{j} &:&  &\cdots \\ E^{0} &:
C_{0}&=&2b_n^2 M .
\end{aligned}
\end{equation}

Let us consider divisor of points $P_{1}=(\mu_{1},w_{1}),
\dots,P_{n}=(\mu_{n},w_{n})$ on $\Gamma$. Substituting (\ref{be})
into (\ref{sgneq}) and setting $E=\mu_1,\dots,\mu_n$ successively,
one gets the following system of equations describing evolution of
the points under the stationary flow,
\begin{equation} \label{mux}
\mu'_{i}\equiv \frac{ \partial \mu_i}{\partial x}=
\frac{\sqrt{R(\mu_i)} } {\mu_i\prod_{j\ne i}^n(\mu_{i}-\mu_{j})}\, ,
\end{equation}
where
\begin{equation} \label{polyfirst}
R(\mu)=\mu C(\mu)= -L^2_0\mu\prod_{r=1}^{2n}(\mu-m_{r}).
\end{equation}
In the case of equation (\ref{SW-sw-eqn}), this should be replaced
by
\begin{equation} \label{polyfirst2}
R(\mu)=\mu\prod_{r=1}^{2n+1}(\mu-m_{r}).
\end{equation}

We now proceed to describe  finite-gap solutions of equation
(\ref{dym-sw-eqn}) and the other equations from its hierarchy.
According to a general theory (see, e.g., Dubrovin [1981],
Belokolos et al. [1994], for any fixed $t$, the $x$-profile of an
$n$-gap solution of an integrable PDE satisfies the $n$-th
stationary equation of the hierarchy. Hence, $n$-gap solutions
$M(x,t_k)$ of $k$-th equation of HD hierarchy must satisfy the
stationary generating equation (\ref{sgneq}) represented by the
Lax pair (\ref{Lax-stationary}), as well as the dynamical
generating equation
\begin{equation} \label{k-generating}
\frac{\partial V}{\partial t_k} =-\frac{1}{2}\frac{\partial^3
B_k}{\partial x^3} +2\frac{\partial B_k}{\partial x} V
+B_k\frac{\partial V}{\partial x} \,, \qquad V=\frac
{M(x,t_k)}{2E} ,
\end{equation}
where the coefficients of $B_k(E)$ are found recursively. Notice
that the latter equation is equivalent to the matrix commutativity
relation
\begin{equation} \label{t-x-Lax}
\left[ \frac{\partial}{\partial x}+{\cal L},
\frac{\partial}{\partial t_k}+W_k\right]=0,
\end{equation}
where
\begin{equation}
W_k(E)=\begin{pmatrix}  -\frac 12 B_k' & B_k \\
                       -\frac 12 B_{k}''+B_k\frac M E  &  \frac 12 B_{k}'
\end{pmatrix} ,
\end{equation}
and $\cal L$ is defined in (\ref{Lax-stationary}). The
compatibility of conditions (\ref{t-x-Lax}),
and (\ref{Lax-stationary}) leads to the following Lax pair
\begin{gather} \label{t_k-Lax}
\frac{\partial}{\partial t_k} W_n(E)=-[W_n(E),W_k(E)] , \qquad
k\in {\mathbb N}, \quad k\ne n,
\end{gather}
For $k=n$, we replace (\ref{t_k-Lax}) with the Lax pair
(\ref{Lax-stationary}) thus identifying $t_n$ with $x$.

The (1,2)-entry of the matrix equation (\ref{t_k-Lax}) implies the
following $t_k$-evolution of the polynomial $B_n(E)$
\begin{equation}
\label{dg} \frac{\partial B_n}{\partial t_k}= \frac{\partial
B_n}{\partial x} B_k - B_n \frac{\partial B_k}{\partial x}, \qquad
k\ne n.
\end{equation}
In case $k=n$ this relation is replaced by $$ \frac{\partial
B}{\partial t_n} = v b_0 \frac{\partial B}{\partial x}\,, $$ where
$v$ is a constant, which can always be eliminated by rescaling
$t_n$.

Expanding the right hand side of (\ref{dg}) in $E$ and using the
condition that it must be a polynomial of degree $n-1$, we find
\begin{equation}
\label{truncate} B_k(E)=\left[ \frac 1{E^{n-k}} B_n(E) \right]_+,
\end{equation}
where $[\; ]_+$ denotes the polynomial part of expression.

As follows from the first equation in (\ref{ch}),
$M=C_{2n}+b_1''$. On the other hand, according to formula
(\ref{be}), $b_1 = -\sum_{i=1}^n \mu_i$. Finally, using the
definition (\ref{MU}) of $M$ in terms of the solution $U$ and
integrating twice with respect to $x$, we obtain
\begin{equation}
\label{b1U} U=\sum_{i=1}^n \mu_i
+\frac{1}{2}\Big(\kappa-C_{2n}\Big) \, x^2+ K_1 x + K_2 \, ,
\end{equation}
where $K_1$ and $K_2$ are constants of integration. If we assume
that all the variables $\mu_i$ are bounded, which is related to
the choice of sign of the leading order coefficient $C_{2n}$, then
$b_1$ is a bounded function of $x$. To find bounded solutions
$U(x,t)$ of the PDE, we set $$ C_{2n}=\kappa, \quad \hbox{and}
\quad K_1=0 \, . $$ Hence, when the above requirements are
imposed, we see that the leading order coefficient of the
polynomial $C(E)$ must coincide with the parameter $\kappa$ of the
PDE.

The Dym equation (\ref{dym-sw-eqn}) is invariant under the
Galilean transformation $$ \hat{x}=x+K_2t, \quad \hat{t}=t,
\quad \hat{U}=U-K_2 \, , $$ so that the constant $K_2$ can
always be eliminated from expression (\ref{b1U}). Therefore, under
the boundedness conditions above, and up to a Galilean
transformation, we assume that the finite-gap and soliton
solutions of the Dym equation (\ref{dym-sw-eqn}) is reconstructed
in terms of the root variables $\mu's$ by the ``trace'' formula
which in case of equations (\ref{dym-sw-eqn}) and
(\ref{SW-sw-eqn}) have the form
\begin{equation}
\label{trc} U(x,t)=\sum_{i=1}^n \mu_i - \mathfrak{m} \, .
\end{equation}
Here $\mathfrak{m}$ is a constant, which equals zero in the case of equation
(\ref{dym-sw-eqn}).

Through~(\ref{trc}) a solution of the system~(\ref{mux}) allows to
construct the instantaneous profile of $U(x,\cdot)$ from a set of
initial conditions $\mu_i(x,\cdot)=\mu_i(0,\cdot)\in
[m_{2i},m_{2i+1}]$, $i=1,\dots,n$. Here the ``dot'' notation
stresses the fact that time $t$ is just a parameter in this
system.

On the other hand, substitution of~(\ref{be}) into (\ref{dg}),
setting $E=\mu_1,\dots,\mu_n$ successively, and taking into
account expressions (\ref{mux}) results in the following
$t_k$-evolution equations for $\mu_i$,
\begin{equation}
\label{higher-t} \frac{\partial\mu_i}{\partial t_k} =B_k(\mu_i)
\frac{\partial\mu_i}{\partial t_x} =B_k(\mu_i)
\frac{\sqrt{R(\mu_i)}} {\mu_i\prod_{j\ne i}^n (\mu_{i}-\mu_{j})},
\quad i=1,\dots, n,
\end{equation}
where, in view of (\ref{be}) and (\ref{truncate}), for
$k=1,\dots,n-1$ $$ B_k(\mu_i)=
\operatorname{Res}\limits_{s=0}\frac 1{s^{n-k}}
\frac{(s-\mu_1)\cdots (s-\mu_n)}{s-\mu_i}, $$ i.e., up to the
sign, the $k$-th elementary symmetric function of $\{\mu_1,\dots
\mu_n\}\setminus \mu_i$. In the case $k=1$,
\begin{equation} \label{mut}
\dot {\mu}_{i}\equiv \frac{ \partial\mu_i}{\partial t_1}
=\frac{(\mu_i-\Sigma)\sqrt{R(\mu_i)}} {\mu_i\prod_{j\ne i}^n
(\mu_{i}-\mu_{j}) }, \;\; i=1,\dots,n, \quad
\Sigma=\mu_1+\cdots+\mu_n ,
\end{equation}
solution of which produces $\mu$'s, and hence the PDE's solution
$U$, at any (later) time $t$.
We notice that for $k>n$, the derivatives $\partial/\partial t_k$
are linear combinations of $\partial/\partial t_1,\dots,\partial/\partial t_n$.

Expressions (\ref{mux}), (\ref{higher-t}), and (\ref{mux}) provide
the so-called {\it $\mu$-variables representation} for the
finite-gap solutions of an evolution equation. They are the
analogs of Drach--Dubrovin equations which describe evolution of
points on the spectral hyperelliptic curve in the case of the KdV
equation. (For further details see Dubrovin [1975], Drach [1919],
Alber {\it et al.}
[1994, 1995, 1999], Gesztesy {\it et al.} [1996], and Alber and
Fedorov [2001].)

With the initial conditions chosen the right-hand-side of system
(\ref{mux}) is real, and the derivative of $\mu_i$ changes sign
when $\mu_i$ reaches the end points of its gap, $\mu_i=m_{2i}$ or
$\mu_i=m_{2i+1}$, corresponding to change of the sheet of the
spectral curve $\Gamma$. Thus each variable $\mu$ undergoes (real)
oscillations between the end points of a gap (so that the
resulting PDE solution $U(x,t)$ remains real).
\medskip

\noindent{\bf Remark.} The condition that the root variables $\mu$'s
are real (or, equivalently, their initial conditions are chosen
as described above), while certainly sufficient
to assure reality of the PDE's solution
$U$ resulting from~(\ref{b1U}), is clearly not necessary (namely, some of
the $\mu$'s could occur in conjugate pairs). A wider class of real
solutions $U$
could be constructed  by relaxing the reality assumption on the
$\mu$-variables.  However, a thorough discussion of the reality
condition for $U$  and its
implications for the root variables, while certainly
desirable, lies beyond the scopes of the present paper,
and it will be addressed in future work.
\medskip

By rearranging and summing up equations (\ref{mux}) and
(\ref{mut}), (\ref{higher-t}), one obtains the following
nonstandard Abel--Jacobi equations
\begin{equation}
\label{A-J} \sum_{i=1}^n \frac{\mu_i^{k}\, d \mu_i}
{\sqrt{R(\mu_i)}} = \left\{ \begin{array}{lll}  d t_{k} &
\mbox{$k=1,\dots,n-1,$} \\
                                  d x & \mbox{$k=n,$} \end{array} \right.
\end{equation}
which contain ($n-1$) holomorphic differentials and one {\it
meromorphic} differential on $\Gamma$. Thus, the number of
holomorphic differentials is less than genus of the Riemann
surface, which implies that the corresponding inversion problem
cannot be solved in terms of meromorphic functions of $x$ and
$t_1,\dots,t_{n-1}$ (see e.g., Markushevich [1977]).

\paragraph{Finite-gap stationary flows in $x$.}
Let us first consider the $x$-flow by fixing time variables in
(\ref{A-J}): $t_k=t_k^0=$const, $k=1,\dots,n-1$, so that $dt_k=0$.
Now introduce a new spatial variable $x_1$ defined as follows
\begin{equation} \label{x-x1}
x=\int_{0}^{x_1}  \frac{1}{L _0}\, \mu_1\cdots\mu_n \, dx_1.
\end{equation}
In view of the well-known Jacobi identities
\begin{equation} \label{Jacobi}
\frac{\mu_i^k}{\prod_{j\ne i}^n (\mu_{i}-\mu_{j})} =\left\{
\begin{array}{lll}   1/(\mu_1\cdots \mu_n) & \mbox{$k=-1,$} \\
   0 & \mbox{$k=0,\dots,n-2,$} \\
           1 & \mbox{$k=n-1,$} \\
\Sigma  & \mbox{$k=n,$} \end{array} \right. ,
\end{equation}
the equations (\ref{mux}) give rise to the following system
\begin{equation} \label{diff1x}
\sum_{i=1}^n \int_{\mu_0}^{\mu_i} \frac{\mu^{k-1}\, d
\mu}{\sqrt{R(\mu)}} = \left\{ \begin{array}{ll} \, x_1+\phi_1 &
\mbox{$k=1,$} \\
                                     \phi_k &
\mbox{$k=2,\dots,n, $}\end{array} \right.
\end{equation}
where $\phi_1,\dots,\phi_n$ are constant phases which depend on
$t_k^0$ as on parameters.

Equations (\ref{diff1x}) include $n$ holomorphic differentials on
$\Gamma$ and determine the standard Abel--Jacobi map of the
symmetric product $\Gamma^{(n)}$ of $n$ copies of $\Gamma$ to the
Jacobi variety (Jacobian) Jac$(\Gamma$). Thus, the flow generated by the
system (\ref{mux}) is made linear on Jac$(\Gamma$) after
introducing the reparametrization (\ref{x-x1}). By using standard
methods (see e.g., Dubrovin [1981] or Mumford [1983]), the map
can be inverted, resulting in expressions for algebraic symmetric
functions of $\mu$-variables in terms of theta-functions of $n$
arguments which depend linearly on $x_1$ and, in a transcendental
way, on $t_k^0$ as parameters. Then, by using the trace formula
(\ref{trc}), one obtains a theta-functional expression for $U$ as
a function of $x_1,t_k^0$, $U=\tilde U(x_1,t_k^0)$.

On the other hand, substituting the theta-functional expression
for the product $\mu_1\cdots\mu_n$  into (\ref{x-x1}) yields a
quadrature. By solving it, one finds $x$ as a meromorphic function
of $x_1$ which depends on $t_0$ as a parameter. However, the
inverse function $x_1(x,t_0)$ is no longer meromorphic in $x$.

Finally, the composition function $U(x,t_0)=\tilde U(x_1(x,t_0),t_k^0)$
gives a profile of the finite-gap solutions of the
(\ref{dym-sw-eqn}) or (\ref{SW-sw-eqn}) equation (for explicit theta-functional
expressions $\tilde U(x_1,t_0)$, $x(x_1,t_0)$ see Alber and Fedorov
[2001]). Notice that as seen from (\ref{x-x1}) and (\ref{diff1x}),
the original $x$-flow is also made linear on Jac$(\Gamma$).
However the straight line motion is not uniform.

The transformation (\ref{x-x1}) involving $x$ and $x_1$ coincides
with a change of variable in the well-known Liouville
transformation  (see, e.g., Verhulst [1996]).

\paragraph{Finite-gap flows in $t_k$.}
Now let us fix the coordinate $x=x_0$ as well as all the times
$t_1,\dots,t_{n-1}$ but $t_k$. Then introduce a new time variable
$\tilde t$ defined by
\begin{equation} \label{t-t1}
dt_k=\frac{\mu_1\cdots\mu_n}{L_0(\Sigma_{k-1})} d\tilde t,
\end{equation}
where $\Sigma_{k-1}$ are the elementary symmetric functions of
$\mu_1,\dots,\mu_n$ such that
\[
(s-\mu_1)\cdots (s-\mu_n)=s^n+s^{n-1}\Sigma_1+\cdots+s^0 \Sigma_n.
\]
Applying again the identities (\ref{Jacobi}), from (\ref{mut}) and
(\ref{higher-t}) we arrive at the following canonical Abel--Jacobi
mapping
\begin{equation} \label{jac}
\sum_{i=1}^n\int_{\mu_0}^{\mu_i} \frac{\mu^{s-1}\,\, d \mu
}{2\sqrt{R(\mu)}}
       =\left\{ \begin{array}{lll} \psi_1 = \tilde t+\phi_1 & \mbox{$s=1,$} \\
                                   \psi_s = \delta_{s,k} t_k+ \phi_s  &
\mbox{$s=2,\dots,n,$}
\end{array}
\right.
\end{equation}
where $\phi_1,\dots,\phi_n$ are constant phases which depend on
$x_0$ and the rest of times $t_l$ as on parameters, and
$\delta_{ij}$ is the Kronecker delta.

As a result of inversion of (\ref{jac}), elementary symmetric
functions of $\mu'$s and therefore the solution of equations
(\ref{dym-sw-eqn}) and (\ref{SW-sw-eqn}) can be found in terms of
theta-functions of $n$ arguments  which depend linearly on
$\psi_s$. This means that the arguments depend linearly  on
$\tilde t$, as well as on the original time $t_k$. However,
$\tilde t$ itself depends on $t_k$ in a nonlinear way.
\medskip

Indeed, to describe the relation between ${\tilde t}$ and $t_k$, we
substitute the theta-functional expressions for the symmetric
functions $\Sigma_n=\mu_1\cdots \mu_n$ and $\Sigma_{k-1}$ into
(\ref{t-t1}). As a result, in contrast to the quadrature
(\ref{x-x1}) relating $x$ and $x_1$, we now get {\it a differential
equation} of the form
$$
\frac{dt_k}{d\tilde t}=F(t_k,\tilde t|x_0),
$$
where $F$ is a transcendental function of $t,{\tilde
t}$ and the parameter $x_0$. It can be shown that the equation
involves a transcendental integral.

\paragraph{Remarks.} \quad \\
1. In contrast to the $x_1$- and $x$-flows considered above, the
flows generated by (\ref{higher-t}) ($t_k$-flows) including
(\ref{mut}), are {\it nonlinear} flows on the Jacobi variety
Jac$(\Gamma$). From the point of view of algebraic geometry, this
phenomenon constitutes the main difference between solutions of
such well known equations as KdV and sine Gordon equations and
equations of (\ref{dym-sw-eqn}) or (\ref{SW-sw-eqn}) type.
\medskip

\noindent 2. The problem of inversion of the full nonstandard Abel
mapping defined by (\ref{A-J}) can be also studied by using a
generalized Jacobian of the curve $\Gamma$. Namely, one has to
extend the mapping by including an extra holomorphic differential
on $\Gamma$ to get a complete set of such differentials. As a
result of this procedure, one gets a flow on nonlinear
subvarieties (strata) of generalized Jacobians. The complete
algebraic geometrical description and explicit formulae are
presented in Alber and Fedorov [2001].

\section{Flows on $n$-dimensional quadrics and
stationary $n$-gap solutions of the (\ref{dym-sw-eqn}) and
(\ref{SW-sw-eqn}) equations.} \setcounter{equation}{0}

Consider a family of confocal quadrics in
$\mathbb{R}^{n+1}=(X_1,\dots,X_{n+1})$
\begin{equation}
\label{confocal} \tilde Q(s)=\left\{\frac{X_1^2}{a_1-s}
+\cdots+\frac{X_{n+1}^2}{a_{n+1}-s}=1 \right\}, \quad
s\in{\mathbb{R}}, \quad 0<a_{n+1}<a_1<\cdots <a_n.
\end{equation}
The elliptic coordinates $\mu_1,\dots,\mu_{n+1}$ can be defined in
$\mathbb{R}^{n+1}$ in a standard way (see, e.g., Jacobi [1884a])
as follows. The condition $s=c$ determines the quadric $\tilde
Q(c)$ on which one of the coordinates, say $\mu_{n+1}$, equals
$c$, and the other coordinates $\mu_1,\dots,\mu_n$ are elliptic
coordinates on $\tilde Q(c)$ defined by relations
\begin{equation} \label{eucl}
X^2_j =(a_j-c) \frac{\prod_{l=1}^{n} (a_j-\mu_l)} {\prod_{k=1,
k\neq j}^{n+1}(a_j-a_k)},  \qquad j=1,\dots,n+1.
\end{equation}
In the sequel without loss of generality we assume $c=0$.

It is well-known that the problem of geodesics on the ellipsoid
$\tilde Q=\tilde Q(0)$ is completely integrable (Jacobi [1884
a,b]). Moreover, as noticed by Jacobi himself and later by many
other authors (see e.g. Rauch-Wojciechowski [1995]), there exists
an infinite sequence of integrable generalizations of the problem
describing a motion on $\tilde Q$ in the force field of certain
polynomial potentials ${\cal V}_p(X_1,\dots,X_{n+1})$,
$p\in\mathbb{N}$ of degree $2p$. The simplest integrable potential is
the quadratic Hooke
potential or the potential of an elastic string joining the center
of the ellipsoid $\tilde Q$ to the point mass on it:
\[
{\cal V}_1= \frac {\sigma}2 (X_1^2+\cdots+X_{n+1}^2), \quad
\sigma= {\rm const}.
\]
In this case in terms of the ellipsoidal coordinates, the total
energy (Hamiltonian) takes the St\"ackel form:
$$
H=\frac 18\sum_{i=1}^{n}\frac{\prod_{j\ne i}
(\mu_i-\mu_j)\mu_i}{\Phi(\mu_i)}
         \left(\frac{d\mu_i}{d\,x}\right)^2
+\frac{\sigma}2 \sum_{i=1}^{n} \mu_i +{\rm const}, $$
where
$$
\Phi(\mu)=(\mu-a_1)\cdots(\mu-a_{n+1}) $$ and $x$ denotes time.
After fixing constants of motion, the system is reduced to the
Abel--Jacobi equations
\begin{eqnarray}
\label{Knorrer} \sum_{k=1}^n \int_{\mu_0}^{\mu_k} \frac{\mu^i\,\,
d \mu}{2\sqrt{ {\cal R}(\mu_k)}} =\delta_{in}\,\,x+\phi_i , \qquad
i=1,\dots,n,  \\ {\cal
R}(\mu)=-\mu\Phi(\mu)[c_0(\mu-c_1)\cdots(\mu-c_{n-1})-\sigma\mu^n],
\quad c _0,\dots,c_{n-1}={\rm const}, \nonumber
\end{eqnarray}
where $\phi_1,\dots,\phi_n$ are constant phases and $c_1,\dots,c_{n-1}$ are
constants of motion.

Notice that for $\sigma=0$ the order of the polynomial ${\cal
R}(\mu)$ is $2n+1$, whereas for $\sigma\ne 0$ it is $2n+2$. The
case $\sigma=0$ corresponds to the free (geodesic) motion on
$\tilde Q$. $c_0$ is the constant in the first integral $(\dot X,\dot X)$
and the remaining constants admit a clear geometric interpretation: the
tangent line to a geodesic is also tangent to the fixed confocal
quadrics $\tilde Q(c_1),\dots,\tilde Q(c_{n-1})$ (Chasles
theorem).

Now notice that equations (\ref{Knorrer}) are equivalent to the
system (\ref{A-J}) with $dt=0$ describing stationary
(\ref{dym-sw-eqn}) and (\ref{SW-sw-eqn}) equations, provided we
identify the roots of the polynomial ${\cal R}(\mu)$ with those of
the odd order polynomial (\ref{polyfirst}) (for $\sigma=0$ and
$L_0=1$) and of the even order polynomial (\ref{polyfirst2}) (for
$\sigma=1$) respectively. The equivalence also holds when some of
the parameters $a_i$ in (\ref{Knorrer}) are negative, which
correspond to the motion on a hyperboloid. For concreteness, we
shall consider only the case of ellipsoids. Taking into account
the trace formula (\ref{trc}), we arrive at the following theorem

\begin{thm}
The geodesic motion and motion in the field of a Hooke potential
on the ellipsoid $\tilde Q$ are linked, at any fixed time $t$, to
the $n$-gap solutions of  (\ref{dym-sw-eqn}) and (\ref{SW-sw-eqn})
equations respectively through the trace formula (\ref{trc}).
Namely, if the roots of the polynomials $R(\mu)$ in
(\ref{polyfirst}) or (\ref{polyfirst2}) coincide with the roots of
${\cal R}(\mu)$ in (\ref{Knorrer}), the profiles of such
solutions are given by the sum of the elliptic coordinates of the
moving point on $\tilde Q$ with addition of ($-\mathfrak{m}$) in
case of equation (\ref{SW-sw-eqn}).
\end{thm}
For the geodesic flow on $\tilde Q$ ($\sigma=0$) and equation
(\ref{dym-sw-eqn}), this result was obtained  in Alber and Alber
[1985], Cewen [1990], and Alber {\it et al.} [1995]).

As with equation (\ref{A-J}), under the change of parameter (\ref{x-x1}),
equations (\ref{Knorrer}) reduce to those containing holomorphic
differentials only and having the same structure as
(\ref{diff1x}). By inverting the corresponding Abel--Jacobi
mapping, one obtains explicit expressions for elementary symmetric
functions of $\mu_i$ and, in view of (\ref{eucl}), for the
Cartesian coordinates $X_1,\dots,X_{n+1}$ in terms of
theta-functions of the new parameter $x_1$ (for the case of the
geodesic flow, see Weierstrass [1844], Moser[1978], and  Kn\"orrer
[1982]). In the case $n=2$, the change of parameter (\ref{x-x1}) was first
applied by Weierstrass [1844] to solve the classical Jacobi
geodesic problem on a triaxial ellipsoid (Jacobi [1884a,1884b]).
\medskip

%%%%%%%%%%%%%%%%%%%%%%%%%%%%%%%%%%%%%%%%%%%%%%%%%%%%%%%%%%%%%%

\section{Billiard dynamical systems and piecewise-smooth weak
solutions of PDE's} \setcounter{equation}{0} In this section it is
first shown how peaked finite-gap solutions of
(\ref{dym-sw-eqn}) and (\ref{SW-sw-eqn}) equations arise in the
limit $m_1\to 0$, where $m_1$ is the smallest root of the
polynomial $R(E)$ in equations (\ref{mux})--(\ref{mut}). Then a
connection to ellipsoidal and hyperbolic billiards is established.

\paragraph{Ellipsoidal billiards and generalized Jacobians.}
Suppose that one of the semi-axes of the ellipsoid $\tilde Q$
tends to zero, namely, $a_{n+1}\to 0$. In the limit, $\tilde Q$
passes into the interior of $(n-1)$-dimensional ellipsoid $$
Q=\{X_1^2/a_1+\cdots+X_n^2/a_n=1\}\in {\mathbb R}^n, \quad
{\mathbb R}^n=(X_1,\dots,X_n). $$ The elliptic coordinates
$\mu_1,\dots,\mu_n$ on $\tilde Q$ transform to elliptic
coordinates {\it in} ${\mathbb R}^n$ giving
\begin{equation} \label{eucl-}
X^2_j=\frac{\prod_{l=1}^{n} (a_j-\mu_l)} {\prod_{k=1, k\neq
j}^{n}(a_j-a_k)}, \qquad j=1,\dots,n,
\end{equation}
which appear as the corresponding limits of (\ref{eucl}).

Then the motion on $\tilde Q$ gets transformed into {\it billiard\/}
motion inside the ellipsoid $Q$. Geodesics on $\tilde Q$ pass into
straight line segments inside $Q$, whereas the points of
intersection of the geodesics with the plane $\{X_{n+1}=0\}$ are
mapped into {\it impact points\/} on $Q$ with elastic reflection.
Also, the motion on $\tilde Q$ under the Hooke force passes to the
motion inside $Q$ under the action of the Hooke force with the
potential ${\cal V}=\sigma(X_1^2+\cdots+X_{n}^2)/2$.  However, in
contrast to cases $\sigma=0$ or $\sigma<0$, for $\sigma>0$ (an
attracting Hooke potential), for the trajectory to reach $Q$ the
total energy $h$ must be sufficiently large. Namely, there ought
to exist a positive $\varepsilon$ such that inside $Q$ the
following double inequality holds
$$
h+\sigma(X_1^2+\cdots+X_{n}^2)/2>\varepsilon
>0. $$
Under this condition, the motion on $\tilde Q$ transforms
to {\it billiard\/} motion inside the ellipsoid $Q$ again having
impacts and elastic reflections along $Q$. Thus, we have ``an
ellipsoidal billiard with the Hooke potential''  which is
described by the mapping ${\cal B}\; :\,({\mathbf   x},{\mathbf
v})\to ({\tilde {\mathbf   x}},
{\tilde {\mathbf   v}})$, where ${\mathbf   x},{\mathbf   v}\in
{\mathbb R}^n$ are the Cartesian
coordinates of a point on $Q$ and the {\it starting} velocity
vector respectively, while $({\tilde {\mathbf   x}}, {\tilde {\mathbf
v}})$ are  the
coordinates and the starting velocity at the next impact point. Following
Fedorov [2001], the mapping has the form
\begin {eqnarray}
\label {map-hooke} {\tilde {\mathbf   x}}&=&\frac {-1}\nu\,
[(\sigma-({\mathbf   v},a^{-1}{\mathbf   v}) ){\mathbf
x}+2({\mathbf   x},a^{-1}{\mathbf   v}){\mathbf   v}], \nonumber \\
{\tilde {\mathbf   v}}&=&\frac {-1}\nu\, [(\sigma-({\mathbf
v},a^{-1}{\mathbf   v}) ){\mathbf   v}-2\sigma ({\mathbf
x},a^{-1}{\mathbf   v}){\mathbf   x}]
+\varrho a^{-1}{\tilde {\mathbf   x}}  \\
&=& \frac {-1}\nu\,
[(\sigma-({\mathbf   v},a^{-1}{\mathbf   v}) )({\mathbf   v}+\varrho
a^{-1}{\mathbf   x}) +2 ({\mathbf   x},a^{-1}{\mathbf   v}) (\varrho
a^{-1} {\mathbf   v}-\sigma {\mathbf   x})],  \nonumber
\end{eqnarray}
$$
\nu=\sqrt{4\sigma ({\mathbf   x},a^{-1}{\mathbf
v})^2+(\sigma-({\mathbf   v},a^{-1}{\mathbf   v}) )^2}, \quad
\varrho=\frac{2({\tilde {\mathbf   v}},a^{-1}{\tilde {\mathbf   x}})}{({\tilde
{\mathbf   x}},a^{-2}{\tilde {\mathbf   x}})}.
$$
Notice that in the limit $\sigma\to 0$
this reduces to a standard billiard mapping given in Veselov
[1988]
$$
{\tilde {\mathbf   x}}={\mathbf   x}-\frac{2({\mathbf
x},a^{-1}{\mathbf   v})}{({\mathbf   v},a^{-1}{\mathbf   v})}\,
{\mathbf   v}, \quad
{\tilde {\mathbf   v}}={\mathbf   v}+\frac{2({\tilde {\mathbf
v}},a^{-1}{\tilde {\mathbf   x}})}{({\tilde {\mathbf
x}},a^{-2}{\tilde {\mathbf   x}})}\, a^{-1}{\tilde {\mathbf   x}}.
$$
The mapping (\ref{map-hooke}), as well as the billiard limits of the motion on
$\tilde Q$ with the higher order potentials
${\cal V}_p(X_1,\dots,X_n,X_{n+1})$ ($X_{n+1}=0$) are completely integrable.

In the limit $a_{n+1}\to 0$ and after using the change of variable
(\ref{x-x1}), the Abel--Jacobi equations (\ref{Knorrer})  are
transformed as follows
\begin{gather}\label{A-J-generalized}
\begin{aligned}
  \sum_{k=1}^n \int_{\mu_0}^{\mu_k}
\frac{\mu^{i-1}\,\, d \mu }{2\sqrt{\rho(\mu)}} &=\phi_i={\rm
const}, \qquad i=1,\dots,n-1,  \\
\sum_{k=1}^n \int_{\mu_0}^{\mu_k} \frac{\, d \mu }
{2\mu\sqrt{\rho(\mu)}}&=x_1+\phi_n, \end{aligned} \rbrace \\
\rho(\mu)=-(\mu-a_1)\cdots(\mu-a_n)\,
[c_0(\mu-c_1)\cdots(\mu-c_{n-1})-\sigma\mu^n] \nonumber
\end{gather}
This system contains $n-1$ holomorphic differentials on the
Riemann surface ${\cal C}=\{w^2=\rho(\mu)\}$ of genus $g=n-1$ and
one differential of the third kind having a pair of simple poles
$\mathcal{Q}_-$, ${\mathcal Q}_+$ on $\cal C$ with $\mu({\mathcal
Q}_\pm)=0$. Here again $\phi_1,\dots,\phi_n$ are constant phases and
$c_0,\dots,c_{n-1}$ are constants of motion. The elliptic
coordinates $\mu_1,\dots,\mu_n$ represent the divisor of $n$
points $P_i=(\mu_i,w_i)$ on ${\cal C}$.

The equations (\ref{A-J-generalized}) describe a well defined
mapping of the symmetric product ${\cal C}^{(g+1)}$ to \\
Jac$({\cal C}, {\mathcal Q}_-,{\mathcal Q}_+)$, the
$(g+1)$-dimensional {\it generalized Jacobian} of the curve
${\cal C}$ with two distinguished points ${\mathcal Q}_\pm$. The later is
obtained from the genus $n$ curve $w^2={\cal R(\mu)}$ in
(\ref{Knorrer}) as a result of confluence of two Weierstrass
points ($a_{n+1}\to 0$) and regularization: cutting out the double
point and gluing $\mathcal{Q}_-, \mathcal{Q}_+$.

The generalized Jacobian is a noncompact algebraic variety which
is topologically equivalent to the product of the customary
$g$-dimensional Jacobian variety Jac$({\cal C})$ with complex
angle coordinates $\phi_1,\dots,\phi_g$ and the cylinder
${\mathbb{C}}^*={\mathbb{C}}\setminus\{0\}$ (for the definition
and description of generalized Jacobians see, among others,
Serre [1959], Previato [1985], Gavrilov [1999], and Fedorov [1999]).

As follows from  (\ref{A-J-generalized}), the geodesic and the
potential billiard motion parameterized by $x_1$ is represented by
a straight line flow on Jac$({\cal C}, {\mathcal Q}_-,{\mathcal
Q}_+)$, which is directed along the real section of ${\mathbb
C}^*$ and leaves the coordinates $\phi$ on Jac$({\cal C})$ invariant.

As we shall see below, for the odd order and even order polynomial
$R(\mu)$ the solutions to the generalized inversion problem
(\ref{A-J-generalized}) have different structures.

\paragraph{Solutions in terms of generalized theta-functions.}

First we concentrate on straight line billiards corresponding to
the case $\sigma=0$ when the curve $\cal C$ has one infinite point
$\infty$. Fix a canonical basis of cycles $A_1,\dots,A_g$,
$B_1,\dots,B_g$ on $\cal C$ and let
$\bar{\omega}_1,\dots,\bar{\omega}_g$ be the dual basis of {\it
normalized} holomorphic differentials on $\cal C$ and
$z_1,\dots,z_g$ be  corresponding coordinates on the universal
covering of Jac$({\cal C})$. There exists a unique $g\times g$
constant normalizing matrix $D$ such that
\begin{equation}
\label{norm} \bar{\omega}_k=\sum_{j=1}^g \frac{D_{kj}\,
\mu^{j-1}\,\, d \mu }{\sqrt{\rho(\mu)}}, \quad z_k=\sum_{j=1}^g
D_{kj}\phi_j, \quad k=1,\dots,g=n-1.
\end{equation}
Let us also introduce a {\it normalized} differential of the third
kind $\Omega_0$ having simple poles at $\mathcal{Q}_\pm$ with
residues $\pm 1$ respectively:
\begin{equation}
\label{Omega_0} \Omega_0=\frac{\sqrt{\rho(0)}\, d \mu
}{\mu\sqrt{\rho(\mu)}} +\sum_{k=1}^g \beta_k\bar{\omega}_k, \qquad
\sqrt{\rho(0)}=\sqrt{a_1\cdots a_n\cdot c_1\cdots c_{n-1}},
\end{equation}
where $\beta_k$ are unique constants such that $\Omega_0$ has zero
$A$-periods on $\cal C$. Then the last equation in
(\ref{A-J-generalized}) can be represented in the following form
\begin{equation}
\label{Zx} \sum_{k=1}^n \int_{\mu_0}^{\mu_k} \Omega_0=Z, \qquad
Z=2\sqrt{\rho(0)}x_1+{\rm const}.
\end{equation}
Notice that in case of the ellipsoidal billiards $\sqrt{R(0)}$ is
always real and hence $Z$ is also real. Let us also choose the
base point $(\mu_0,w_0)$ of the mapping (\ref{A-J-generalized}) to
be an infinite point $\infty\in{\cal C}$. According to Fedorov
[1999], the solution of the problem of inversion
(\ref{A-J-generalized}) together with (\ref{eucl-}) yields the
following expressions for the Cartesian coordinates $X_i$ of the
point moving inside the ellipsoid $Q$:
\begin{gather}
\label{segment}
X_i(x_1,z)=\kappa_i\, \frac{e^{-Z/2}{\theta}[\Delta+\eta_{(i)}](z-q/2)+
e^{Z/2}{\theta}[\Delta+\eta_{(i)}](z+q/2)}
{e^{-Z/2}{\theta}[\Delta](z-q/2)+ e^{Z/2}{\theta}[\Delta](z+q/2)},
\quad i=1,\dots,n,  \\
z=(z_1,\dots,z_{n-1})^T,\quad
Z=2\sqrt{R(0)}x_1+Z_{0}, \quad z,Z_{0} ={\rm const}, \nonumber \\
q=2\left(\int_{\infty}^{\mathcal{Q}_+}\bar{\omega}_1,\, \dots,
\int_{\infty}^{\mathcal{Q}_+}\bar{\omega}_g\right)^T\in
{\mathbb{C}}^g, \quad \kappa_i={\rm const}.     \nonumber
\end{gather}
These expressions involve quotients of generalized
theta-functions, where ${\theta}[\Delta+\eta_{(i)}](z)$ and
${\theta}[\Delta](z)$ are customary theta-functions associated
with the Riemann surface ${\cal C}$ with appropriately chosen
half-integer theta-characteristics $\eta_{(i)}$ ($\Delta$ is the half-integer
theta-characteristic corresponding to the vector of Riemann's constants).
The vector $q$ coincides with the vector of $B$-periods of the
meromorphic differential $\Omega_0$. The constant factors $\kappa_i$
depend on the parameters of the curve $\cal C$ only.
(For the definition and properties of the generalized
theta-functions see e.g., Belokolos {\it et al.} [1994], Gagnon
{\it et al} [1992], Ercolani [1987], and Fedorov [1999].)

The expressions (\ref{segment})  describe a straight line segment
in $\mathbb{R}^n$ ($\mathbb{C}^n$) with $z$ playing a role of a
constant phase vector which defines the position of the segment.
When one of the $\mu$-variables, say $\mu_1$, equals zero, the
corresponding point $P_1=(\mu_1,\sqrt{R(\mu_1)})$ on the curve
$\cal C$ coincides with one of the poles $\mathcal{Q}_-$,
$\mathcal{Q}_+$ of the differential $\Omega_0$. Then, as follows
from the mapping (\ref{A-J-generalized}) and (\ref{Zx}), $x_1$ and
$Z$ become infinite. On the other hand, in view of (\ref{eucl-}),
at this moment the moving point in ${\mathbb{R}}^n$ meets an ellipsoid $Q$.

It follows that as $x_1$ and $Z$ change from $-\infty$ to $\infty$
along the real axis, the expressions (\ref{segment}) have finite
limits, giving the coordinates of two subsequent impact points on
$Q$. Notice that $X_i(\infty,z)$ have the same values as
$X_i(-\infty,z+q)$. Hence the next segment of the billiard
trajectory is given by (\ref{segment}) with $z$ being replaced by
$z+q$. This yields the following algebraic-geometrical description
of the billiard motion (see also Fedorov [1999]).

\begin{thm} As the point mass inside $Q$  approaches the ellipsoid,
the point $P_1$ on $\cal C$ tends to the pole $\mathcal{Q}_+$. At the
moment of impact, $P_1$ jumps from $\mathcal{Q}_+$ back to
$\mathcal{Q}_-$, whereas the phase vector $z$ is increased by $q$.
The process repeats itself  for each impact.
\end{thm}
Using this property and by applying induction, from
(\ref{segment}) the coordinates of the whole sequence of impact
points are found in the form
\begin{equation}
\label{impact} {\mathbf   x}_i(N)=\kappa_i\frac{
{\theta}[\Delta+\eta_{(i)}](z_0+Nq) } {{\theta}[\Delta](z_0+Nq)},
\qquad i=1,\dots,n
\end{equation}
where $N\in{\mathbb{N}}$ is the number of impacts and the phase vector
$z_0=(z_{10},\dots,z_{g0})^T$ is the same for all the segments of
the billiard trajectory.

These expressions depend on customary theta-functions only and, as
functions of $z_0$, are meromorphic on a covering of the Jacobian
variety of $\cal C$. They have also been obtained by Veselov
[1988] by using a factorization of matrix polynomials (see also
Moser and Veselov [1991]). The work of Veselov is closely related
to the discretization of mechanics that preserves the integrable
structure. The numerical implementation of Veselov's procedures
was given in Wendlandt and Marsden [1997], a discrete reduction
procedure in Marsden, Pekarsky and Shkoller [1999], Bobenko and
Suris [1999] and an extension to PDE's in Marsden, Patrick and
Shkoller [1999].

The generalized Abel map (\ref{A-J-generalized}) yields
expressions in terms of generalized theta-functions for the
elementary symmetric functions of the variables $\mu$. In
particular, following Fedorov [1999],  one obtains
\begin{align}
\label{prod_mu} \mu_1\cdots\mu_n &
=\partial_{x_1}\partial_V\log\tilde{\theta}[\Delta](z,Z) \nonumber
\\
            & =2\sqrt{\rho(0)}\,\partial_Z\,
\frac{e^{-Z/2}\partial_V{\theta}[\Delta](z-q/2)
+e^{Z/2}\partial_V{\theta}[\Delta](z+q/2) }
{e^{-Z/2}{\theta}[\Delta](z-q/2)+e^{Z/2}{\theta}[\Delta](z+q/2)}\, ,
\end{align}
where
\begin{eqnarray}
\label{gen-theta}
\tilde\theta[\Delta](z,Z)=e^{-Z/2}{\theta}[\Delta](z-q/2)
+e^{Z/2}{\theta}[\Delta](z+q/2), \nonumber \\
Z=2\sqrt{\rho(0)}x_1+Z_{0}, \qquad
\partial_V=V_1\frac{\partial}{\partial z_1}+\cdots+
V_n\frac{\partial}{\partial z_n},
\end{eqnarray}
and where $V$ is the last column of the normalizing matrix $D$
defined in (\ref{norm}): $V=(D_{1g},\dots,D_{gg})^T$. The phases
$z$ and $Z_0$ are the same as in (\ref{segment}). As follows from
(\ref{prod_mu}), for $x_1,Z\to\pm\infty$, the product
$\mu_1\cdots\mu_n$ tends to zero, as expected. Taking the integral
(\ref{x-x1}) with $L_0=1$ yields
\begin{align}
\label{x-x} x(x_1,z) & =\int \mu_1\cdots\mu_n\, dx_1=
\partial_V \log \tilde{\theta}(z,Z)+{\rm const} \nonumber\\
            & = \frac{e^{-Z/2}\partial_V{\theta}[\Delta](z-q/2)
+e^{Z/2}\partial_V{\theta}[\Delta](z+q/2) }
{e^{-Z/2}{\theta}[\Delta](z-q/2)+e^{Z/2}{\theta}[\Delta](z+q/2)}+{\rm
const}.
\end{align}
It follows from this expression that the original parameter $x$
has finite values as $x_1\to \pm\infty$ and $x(\infty,z)$ has the
same value as $x(-\infty,z+q)$.  Now, substituting in (\ref{x-x})
$Z=-\infty$, $Z=\infty$,  by induction, we find the length of the
$N$-th segment of the billiard trajectory in the form
\begin{equation}
\label{lenght} x(N)-x(N-1)=\frac{\partial_V \theta
[\Delta](z_0+Nq)} {\theta[\Delta](z_0+Nq)} -\frac{\partial_V
\theta [\Delta](z_0+Nq-q)} {\theta[\Delta](z_0+Nq-q)}, \qquad N\in
{\mathbb{N}} .
\end{equation}
$z_0$ being the same as in (\ref{impact}).
\medskip

As a result, the solution $X_i(x)$, $x\in{\mathbb R}$, of the {\it
continuous\/}
geodesic billiard problem should be viewed as consisting of
infinite number of pieces each parameterized by $x_1\in
(-\infty,\infty)$ and given by (\ref{segment}) and (\ref{x-x}).
These pieces are obtained by iteratively adding vector $q$ to the
phase $z$ in (\ref{segment}) and (\ref{x-x})  and they are glued
together at the impact points corresponding to $x_1=\pm\infty$.

\medskip

Now we turn to the ellipsoidal billiard with the Hooke potential
($\sigma=1$). In this case the curve $\cal C$ appearing in
(\ref{A-J-generalized}) has 2 infinite points at $\pm \infty$. We
again introduce normalized differentials $\bar{\omega}_k$,
$\Omega_0$, and coordinates $z_k$, $Z$ according to (\ref{norm})
and (\ref{Zx}). Let the base point of the mapping
(\ref{A-J-generalized}) be one of the Weierstrass points of $\cal
C$, say $\mu_0=a_n$. Then, instead of (\ref{segment}), the
inversion of the generalized mapping (\ref{A-J-generalized})
yields the following expressions for the {\it squares} of the
Cartesian coordinates of the mass point moving inside an ellipsoid
$Q$:
\begin{eqnarray}
\label{x-squared} X_i^2(x_1,z)=\kappa_i'\, \frac{
{\tilde\theta}^2[\Delta+\eta_{(i)}](z,Z) } {
\tilde{\theta}[\Delta] (z-{\hat q}/2,Z-{\hat S}/2)\,
        \tilde{\theta}[\Delta](z+{\hat q}/2,Z+{\hat S}/2) }, \quad i=1,\dots,n,
        \nonumber \\
z=(z_1,\dots,z_{n-1})^T,\quad Z=2\sqrt{\rho(0)}x_1+Z_{0}, \quad
z,Z_{0} ={\rm const}, \nonumber \\
q=\int_{\mathcal{Q}_-}^{\mathcal{Q}_+}\,(\bar{\omega}_1,\dots,\bar{\omega}_g)^T,
\quad {\hat
q}=2\int_{a_n}^{\infty_+}\,(\bar{\omega}_1,\dots,\bar{\omega}_g)^T,
\quad {\hat S}=\int_{\infty_-}^{\infty_+}\,\Omega_0,
\end{eqnarray}
where $\tilde\theta[\Delta](z,Z)$ is defined in (\ref{gen-theta})
and
\begin{equation}
\tilde{\theta}[\Delta+\eta_{(i)}](z,Z)
=e^{-Z/2}{\theta}[\Delta+\eta_{(i)}](z-q/2)
+e^{Z/2}{\theta}[\Delta+\eta_{(i)}](z+q/2)
\end{equation}
Here $\kappa_i'$ are constants, and $\sqrt{\rho(0)}$ is the same
as in (\ref{Omega_0}). Similarly to (\ref{segment}), as $x_1$ and
$Z$ pass from $-\infty$ to $\infty$, $X_i^2(x_1,z)$ tend to finite
values resulting in the squares of the coordinates of subsequent
impact points on $Q$. Thus, expressions (\ref{x-squared}) describe
a segment of trajectory of the billiard in the field of the Hooke
potential between two impacts. After each impact the phase vector
$z$ changes according to Theorem 4.1. Then, by using induction,
the sequence of impact points is described as follows
\begin{eqnarray}
\label{impact-2} {\mathbf   x}_i^2(N)=\kappa_i'\,\frac{
{\theta}^2[\Delta+\eta_{(i)}](z_0+Nq) } { \theta[\Delta](z_0-{\hat
q}+Nq)\, \theta[\Delta](z_0+{\hat q}+Nq) }, \qquad i=1,\dots,n,
\nonumber\\ N\in{\mathbb{N}}, \quad
z_0=(z_{10},\dots,z_{g0})^T={\rm const}. \qquad
\end{eqnarray}
Apparently, this theta-functional solution for the  billiard with
the Hooke potential was not previously known. Lastly, we find the
following expression for $x$
\begin{equation}
\label{x-x-2} x(x_1,z)={\rm const} +\log\frac{
\tilde{\theta}[\Delta](z-{\hat q}/2,Z-{\hat S}/2) } {
\tilde{\theta}[\Delta](z+{\hat q}/2, Z+{\hat S}/2) }, \quad
Z=2\sqrt{\rho(0)}x_1+{\rm const},
\end{equation}
which, for $x_1\to \pm\infty$ and $Z\to \pm\infty$, has finite
limits determining $x$ for two subsequent impacts. Then, using the
expression (\ref{gen-theta}), by induction, we express a
$x$-interval between the impacts in terms of the customary
theta-function:
\begin{equation}
\label{time-N} x(N)-x(N-1)=\log \frac{\theta[\Delta](z_0-{\hat
q}/2+Nq) } {\theta[\Delta](z_0+{\hat q}/2+Nq) } -\log
\frac{\theta[\Delta](z_0-{\hat q}/2+Nq-q) }
{\theta[\Delta](z_0+{\hat q}/2+Nq-q) } -\log {\hat S}.
\end{equation}
We emphasize that, in contrast to the geodesic billiard, for the
billiard in the potential field the ``time'' $x$ is not
proportional to the length of a trajectory.

\paragraph{Stationary finite-gap peaked solutions.}
Now we return to the finite-gap solutions of equations (HD) and
(SW). Notice that under the limit $m_1\to 0$ the mapping
(\ref{diff1x}) takes the form (\ref{A-J-generalized}) with
$\rho(\mu)$ being a polynomial of degree $2n-1$ and $2n$
respectively.

The trace formula (\ref{trc}) and relations (\ref{eucl-}) yield $$
U=\sum_{j=1}^n X_j^2 +\sum_{i=1}^n a_i+{\mathfrak m}. $$ Then
solution to the billiard problems (\ref{segment})--(\ref{time-N})
provide solutions $U(x,t_0)$ for the above equations which consist
of infinite sequences of smooth pieces each one corresponding to a
segment between two impacts. The impacts themselves give peaks of
$U(x,t_0)$. This leads to the following theorem.

\begin{thm} \label{th2}

$1)$. At any fixed time $t=t_0$, finite-gap peaked solution of the
equation (\ref{dym-sw-eqn}) consists of an infinite number of
pieces $U_N(x,t_0)$, $N\in{\mathbb Z}$ glued at peak points. Let
$\rho(\mu)$ be any polynomial with distinct  roots
$a_1,\dots,a_n$. Then, for any $N$, every piece is given by the
following pair of theta-functional expressions parameterized by
$x_1\in{\mathbb R}$,
\begin{align}
\label{peak1} U_N &=\sum_{j=1}^n X_j^2(x_1,z_N)+\sum_{i=1}^n a_i,
\\
x(x_1,z) & =\frac{e^{-Z/2}\partial_V{\theta}[\Delta](z_N-q/2)
+e^{Z/2}\partial_V{\theta}[\Delta](z_N+q/2) }
{e^{-Z/2}{\theta}[\Delta](z_N-q/2)+e^{Z/2}{\theta}[\Delta](z_N+q/2)}+x_0,
\\ z_N=z_0&+Nq\in{\mathbb C}^{n-1}, \quad Z=2\sqrt{\rho(0)}x_1+Z_0, \nonumber
\end{align}
where $X_j^2(x_1,z)$ and $q$ are given by (\ref{segment}) and $z_0,
Z_0, x_0$ are constant phases of the solution depending on $t_0$,
which are the same for any piece. The length of the $N$-th piece
equals
\begin{equation}
\frac{\partial_V \theta [\Delta](z_0+Nq)} {\theta[\Delta](z_0+Nq)}
-\frac{\partial_V \theta [\Delta](z_0+Nq-q)}
{\theta[\Delta](z_0+Nq-q)} .
\end{equation}
\medskip

$2)$. At any fixed time $t=t_0$ finite-gap peaked solution to
equation (\ref{SW-sw-eqn}) consists of an infinite number of
pieces $U_N(x,t_0)$, $N\in{\mathbb Z}$ which are glued at peak
points. The pieces are given in the following parametric form
\begin{align}
U_N &=\sum_{j=1}^n X_j^2(x_1,z_N)+\sum_{i=1}^n a_i+{\mathfrak m},
\qquad z_N=z_0+Nq\in{\mathbb C}^{n-1}, \\ x(x_1,t_0)& =\log\frac{
\tilde{\theta}[\Delta](z_N-{\hat q}/2,Z-{\hat S}/2) }
{\tilde{\theta}[\Delta](z_N+{\hat q}/2,Z+{\hat S}/2) }+x_0, \quad
Z=2\sqrt{\rho(0)}x_1+Z_0,
\end{align}
where $X_j^2(x_1,z)$ are given by (\ref{x-squared}) and $z_0, Z_0,
x_0$ are constant phases which depend on $t_0$. The $x$-length of
$N$-th piece equals
\begin{equation}
\log \frac{\theta[\Delta](z_0-{\hat q}/2+Nq) }
{\theta[\Delta](z_0+{\hat q}/2+Nq) } - \log
\frac{\theta[\Delta](z_0-{\hat q}/2+Nq-q) }
{\theta[\Delta](z_0+{\hat q}/2+Nq-q) } -\log {\hat S}.
\end{equation}
\end{thm}
\medskip

When in the polynomials (\ref{polyfirst}) or (\ref{polyfirst2})
$m_1=0$ and $m_2$ tends to zero, the distance between subsequent
peaks of a profile tends to zero  and in the limit the peaks
coalesce. (Notice that this is done for a fixed $t$.) The solution
$U(x,t_0)$ for this limiting case is smooth.
\medskip

\noindent{\bf Remark.}
It is known (see, for instance, Fedorov [1999])
that there are special degenerate umbilic billiard solutions of
the classical billiard problem (without a potential) that have
straight line segments meeting $n-1$ fixed focal conics of $Q$
between any subsequent impacts and, as $x\to\pm\infty$, the
billiard motion converges to simple oscillations along the largest
axis of the ellipsoid. This corresponds to the confluence of the
roots of the polynomial $\rho(\mu)$ in (\ref{A-J-generalized}), $$
c_1=a_1, \quad \dots, \quad c_{n-1}=a_{n-1}. $$ As a result, the
hyperelliptic curve $\cal C$ becomes singular of arithmetic genus
zero and the asymptotic billiard motion is described in terms of
tau-functions. The corresponding asymptotic peaked solutions of
equations (HD) and (SW) are given in Alber and Fedorov [2001].

\paragraph{Time-dependent piecewise-meromorphic solutions.}
Now we pass to global algebraic geometrical description of the
finite-gap peaked solutions. After setting  $m_1\to 0$, the
system (\ref{A-J}) is formally reduced to the following Abel--Jacobi mapping
\begin{equation}
\label{generalized-II}
        \int_{\mu_0}^{\mu_1}\, \frac{\mu^{k-1}\,\,
d \mu }{2\sqrt{\rho(\mu)}}+\cdots +
        \int_{\mu_0}^{\mu_{n}}\, \frac{\mu^{k-1}\,\,
d \mu }{2\sqrt{\rho(\mu)}} = \left\{ \begin{array}{lll}
t_{k}+\phi_k  &  \mbox{$k=1,\dots,n-1,$} \\
               x+\phi_n &  \mbox{$k=n,$} \end{array} \right.
\end{equation}
where
$$
\rho(\mu)=-L_0^2\prod_{r=2}^{2n}(\mu-m_r) \quad {\rm and}
\quad \rho(\mu)=\prod_{r=2}^{2n+1}(\mu-m_r)
$$
in case of equations (\ref{dym-sw-eqn}) or (\ref{SW-sw-eqn}) respectively.
Here $\phi_1,\dots,\phi_n$ are constant phases. This system
contains $n-1$ independent holomorphic differentials defined on
the genus $g=n-1$ Riemann surface $\{w^2=\rho(\mu)\}$, which can
be identified with the curve $\cal C$ described above. However, in
contrast to the system (\ref{A-J-generalized}), in  case of a
polynomial $\rho(\mu)$ of odd order which corresponds to equation
(\ref{dym-sw-eqn}), the last equation in (\ref{generalized-II})
contains a meromorphic differential of the {\it second} kind
having a double pole at the infinite point $\infty$ on $\cal C$.
In case of a polynomial $\rho(\mu)$ of even order corresponding to
equation (\ref{SW-sw-eqn}), the last equation includes a
meromorphic differential of the {\it third} kind with a pair of
simple poles at the infinite points $\infty_-, \infty_+$ on $\cal
C$.

According to Clebsch and Gordon [1866] and Gavrilov [1999], in the
odd order case, such a system describes a well defined and
invertible mapping of the symmetric product ${\cal C}^{(g+1)}$ to
Jac$({\cal C},\infty)$, the generalized Jacobian of the curve
${\cal C}$ with {\it one} distinguished point at $\infty$. The set
Jac$({\cal C},\infty)$ is a noncompact algebraic variety which is
topologically equivalent to the product Jac$({\cal C})\times
{\mathbb{C}}$. To describe this case we introduce a normalized
differential of second kind having a double pole at $\infty$,
\begin{equation}
\label{norm2} \Omega_{\infty}^{(1)}=\frac{\sqrt{-1} L_0\mu^{g}\,\, d \mu}
{2\sqrt{\rho(\mu)}} +\sum_{k=1}^{g} d_k\bar{\omega}_k, \qquad
g=n-1,
\end{equation}
where $\bar{\omega}_k$ are the normalized holomorphic
differentials specified in (\ref{norm}), $d_k$ are normalizing
constants such that all $A$-periods of $\Omega_{\infty}^{(1)}$ on
$\cal C$ are zeros. Then the last equation in
(\ref{generalized-II}) implies that
\begin{eqnarray}
\label{Z-norm2} \sum_{i=1}^n \int_{\mu_0}^{\mu_i}
\Omega_{\infty}^{(1)}=Z, \qquad Z=\sqrt{-1}  L_0x +(d,Dt)+{\rm const}, \\
d=(d_1,\dots,d_{n-1})^T, \quad  t=(t_n,\dots,t_2)^T, \nonumber
\end{eqnarray}
where $D$ is an $(n-1)\times (n-1)$ normalizing matrix defined in
(\ref{x-x}).

Since $\infty$ now is a pole of $\Omega_{\infty}^{(1)}$, we choose
the basepoint $P_0=(\mu_0,w_0)$ to be a {\it finite} Weierstrass
point on $\cal C$. For concreteness we choose $P_0=(m_{2n},0)$.
Applying the residue theorem to the generalized theta-function
associated with Jac$(\cal C,\infty)$ we solve the inversion
problem (\ref{generalized-II}) and find the following expression
\begin{eqnarray}
\label{rational}
\sum_{i=1}^n\mu_i=C_1-Z^2+\frac{2Z\partial_V\theta[\Delta+\eta_{2n}](z)
-\partial_V^2\theta[\Delta+\eta_{2n}](z)}
{\theta[\Delta+\eta_{2n}](z)}, \\ Z=\sqrt{-1} L_0x +(d,Dt)+Z_0 , \quad
z=Dt+z_0\in{\mathbb{C}}^{n-1}, \nonumber \\ Z_0, z_0={\rm const},
\qquad C_1=\sum_{k=1}^g\oint_{A_k}\mu\,{\bar\omega_k}+m_{2n},
\nonumber
\end{eqnarray}
where the half-integer characteristic $\eta_{2n}$ labels the point
$(m_{2n},0)$, the vector $V=(D_{1g},\dots,D_{gg})^T$ is specified
in (\ref{norm}), and the constant $C_1$ contains the sum of
integrals along the canonical cycles $A_1,\dots,A_g$ on $\cal C$.
Notice that in the above formula $\partial_V=\partial_{t_2}$.
Expression (\ref{rational}) is meromorphic in $x$ and
$t_1,\dots,t_{n-1}$ and can be regarded as a generalization of the
Matveev--Its formula to the case of the noncompact variety
Jac$({\cal C},\infty)$.
\medskip

In the case of an even order curve $\cal C$, corresponding to
finite-gap peaked solutions of equation (\ref{SW-sw-eqn}),
system (\ref{generalized-II}) defines a mapping of the symmetric
product ${\cal C}^{(g+1)}$ to the generalized Jacobian Jac$({\cal
C},\infty_\pm)$ which is topologically equivalent to the product
Jac$({\cal C})\times {\mathbb{C}}^*$. As above, we set $P_0$ to be
the {\it last} Weierstrass point $(m_{2n+1},0)$ and introduce the
normalized differential of the third kind having a pair of simple
poles at $\infty_-, \infty_+\in {\cal C}$, as well as the
corresponding variable $Z$:
\begin{equation}
\label{norm3} \Omega_{\infty_\pm} =\frac{\mu^{g}\,\, d
\mu}{2\sqrt{\rho(\mu)}} +\sum_{k=1}^{g} {\bar d}_k\bar{\omega}_k,
\qquad Z=\sum_{i=1}^n \int_{m_{2n+1}}^{\mu_i}\,
\Omega_{\infty_\pm},
\end{equation}
where $({\bar d}_1,\dots,{\bar d}_g)={\bar d}$ are chosen such
that all the $A$-periods of $\Omega_{\infty_\pm}$ are zeros. Then,
applying the residue theorem to the generalized theta-function
associated with the Jac$({\cal C},\infty_\pm)$  yields
\begin{equation}
\label{exponential} \sum_{i=1}^n\mu_i+{\mathfrak m} ={\rm const}
-\frac{e^{-Z}\theta[\Delta](z-{\hat q})
+e^{Z}\theta[\Delta](z+{\hat q})}{\theta[\Delta](z)},
\end{equation}
where, in view of (\ref{norm3}),
\begin{align}
\label{Z-norm-3} Z=x+({\bar d},Dt)+Z_0 , \quad
z=Dt+z_0\in{\mathbb{C}}^{g},  \nonumber\\ {\hat
q}=\left(\int^{\infty_+}_{\infty_-}\bar{\omega}_1, \dots,
\int^{\infty_+}_{\infty_-} \bar{\omega}_g\right)^T\in
{\mathbb{C}}^g, \qquad Z_0, z_0={\rm const}.
\end{align}
%Expression (\ref{exponential}) is a generalization of the trace
%formula in case of a polynomial of even order.
\medskip

\noindent {\bf Remark.}
According to the formula (\ref{trc}), expressions (\ref{rational})
and (\ref{exponential}) describe formal solutions to equations
(\ref{dym-sw-eqn}) and (\ref{SW-sw-eqn}) respectively. However,
while treating these solutions, one needs to take into account the
reflection phenomenon described in Theorem 4.1. Namely, when a
certain variable $\mu_i$ passes zero, the point
$P_i=(\mu_i,\sqrt{\rho(\mu_i)})$ jumps from one sheet of the
Riemann surface $\cal C$ to another or, in other words, from the
pole $\mathcal{Q}_+$ of the differential of the third kind
$\Omega_0$ to another pole $\mathcal{Q}_-$. Therefore, the above
expressions do not provide global solutions to the equations.
Instead, the following theorem holds.

\begin{thm}
$1)$. The time-dependent finite-gap peaked solution $U(x,t)$
of (\ref{dym-sw-eqn}) consists of an infinite number of pieces in
${\mathbb R}^n=(t_1,\dots,t_{n-1},x)$ described by meromorphic functions
\begin{eqnarray}
\label{profile-2}
U_N(x,t)=C_1-Z_N^2+\frac{2Z_N\partial_V\theta[\Delta+\eta_{2n}](z_N)
-\partial_V^2\theta[\Delta+\eta_{2n}](z_N)}
{\theta[\Delta+\eta_{2n}](z_N)}, \qquad N\in {\mathbb{Z}},
\nonumber \\
z_N=D{\mathbf t}+Nq+z_0, \quad Z_N=\sqrt{-1} L_0x+(d,z_N)+Nh+Z_0, \qquad
Z_0,z_0={\rm const}, \nonumber \\
{\mathbf t}=(t_1,\dots,t_{n-1}), \quad
h=\int_{\mathcal{Q}_-}^{\mathcal{Q}_+}\,\Omega_{\infty}^{(1)},
\quad q=\left(\int^{\mathcal{Q}_+}_{\mathcal{Q}_-}\bar{\omega}_1,
\dots, \int^{\mathcal{Q}_+}_{\mathcal{Q}_-}\bar{\omega}_g\right)^T,
\end{eqnarray}
where $C_1$ is the constant specified in (\ref{rational}).

For a fixed $N$ the corresponding piece $U_N(x,t)$ is bounded by
nonintersecting
surfaces $\mathcal{S}_{N-1}$ and $\mathcal{S}_{N}$ in
${\mathbb R}^n$ given by equations
\begin{equation}
\label{peak-surface} \mathcal{S}_N=\{x=p_N(t)\}, \quad
p_N(t)=\frac 1{\sqrt{-1} L_0} \left(
\partial_V\log\theta[\Delta+\eta_{2n}](z_N+q/2)-(d,z_N)-Nh
\right).
\end{equation}
The adjacent pieces $U_{N}(x,t)$ and $U_{N+1}(x,t)$ are thus glued
to each other along $\mathcal{S}_N$, where
\begin{equation}
\label{peak-value} U(p_N(t),t)= C_1-\partial_V^2\log
\theta[\Delta+\eta_{2n}](z_N+q/2).
\end{equation}

$2)$. The finite-gap peaked solution $U(x,t)$ of
(\ref{SW-sw-eqn}) consists of an infinite number of pieces given
by meromorphic functions
\begin{eqnarray}
\label{profile-3} U_N(x,t)={\rm
const}-\frac{e^{-Z_N}\theta[\Delta](z_N-{\hat q})
+e^{Z_N}\theta[\Delta](z_N+{\hat q})}{\theta[\Delta](z_N)}, \quad
N\in{\mathbb Z}, \\ z_N=Dt+qN+z_0, \quad Z_N=x+({\bar
d},z_N)+N{\bar h}+Z_0 , \nonumber \\ t=(t_n,\dots,t_2), \quad Z_0,
z_0={\rm const}, \quad
{\bar h}=\int_{\mathcal{Q}_-}^{\mathcal{Q}_+}\,\Omega_{\infty_\pm},
\nonumber
\end{eqnarray}
where the vector $\hat q$ is described in (\ref{Z-norm-3}). The
piece $U_N(x,t)$ is bounded by peak surfaces
$\bar\mathcal{S}_{N-1}$ and $\bar{\mathcal S}_{N}$ defined as follows
\begin{equation}
\label{odd-peak-surface} \bar \mathcal{S}_N=\{ x={\bar p}_N(t)\},
\qquad {\bar p}_N(t)={\rm const}-\log \frac
{\theta[\Delta](z_N-{\hat q}+q/2)}
    {\theta[\Delta](z_N+{\hat q}+q/2)}.
\end{equation}
The adjacent pieces $U_{N}(x,t)$ and $U_{N+1}(x,t)$ are glued
together along $\bar{\mathcal S}_N$, where
\begin{equation}
\label{odd-peak-value} U(p_N(t),t)={\rm const}-\partial_V \log
\frac {\theta[\Delta](z_N-{\hat q})} {\theta[\Delta](z_N+{\hat q})}.
\end{equation}
\end{thm}

Notice that along the peak surfaces, the solutions described in 1)
and 2) have discontinuous partial derivatives with respect to $x$
and $t_1,\dots,t_{n-1}$.
\medskip

\noindent{\bf Remark.}
By fixing all the times but $t_k$ in the above expressions, one
obtains 2-dimensional piecewise solutions $U_N(x,t_k)$, whereas the
corresponding sections of $\mathcal{S}_N, \bar{\mathcal S}_N\subset
(x,t)={\mathbb R}^n$ describe peak lines in $(x,t_k)$-plane. As
follows from (\ref{peak-surface}) and
(\ref{odd-peak-surface}), the motion of the $N$-th peak $p_N(t_k)$
along $x$-axis
is described by a sum of a linear function in $t_k$ and a quasi-periodic one.
The latter function becomes periodic in the case $g=1$.

Finally, after fixing all the times without exception, expressions
(\ref{profile-2}) and (\ref{profile-3}) provide pieces of the
stationary finite-gap peaked solution already described in Theorem 4.2.

\bigskip

\noindent{\it Proof of Theorem\/} 4.3. According to Theorem 4.2, the
profiles of finite-gap peaked solutions are associated with
geodesic ellipsoidal billiards and billiards in the field of a
Hooke potential. An impact point on the boundary of a billiard
trajectory corresponds to a peak of the profile $U(x,t_0)$, and
this happens when one of the $\mu_i$ passes zero. Hence, the
solution (\ref{rational}) is valid until one of the points
$P_1,\dots, P_n$ on $\cal C$ coincides with $\mathcal{Q}_-$ or
$\mathcal{Q}_+$, the poles of the differential $\Omega_0$ in
(\ref{Omega_0}). Putting, for example, $P_n\equiv \mathcal{Q}_+$
($\mu_n\equiv 0$) in (\ref{generalized-II}), one arrives at the
following relations involving the normalized differentials defined
in (\ref{norm}) and (\ref{norm2}),
\begin{align}
\label{good} \sum_{i=1}^{g}  \int_{P_0}^{\mu_i}\, \bar{\omega}_k
&=z_k- q_k/2 , \qquad k=1,\dots,g, \\ \label{2n-kind}
\sum_{i=1}^{g}  \int_{P_0}^{\mu_i}\,
\left(\Omega_{\infty}^{(1)}-\sum_{k=1}^{g}
d_k\bar{\omega}_k\right) & =\sqrt{-1} L_0 x - \int_{P_0}^{\mathcal{Q}_+}
\left(\Omega_{\infty}^{(1)}-\sum_{k=1}^{g}
d_k\bar{\omega}_k\right),
\end{align}
where  $P_0=(m_{2n},0)$. Notice that equations (\ref{good}) form a
closed system for the variables $\mu_1,\dots,\mu_{n-1}$ and
describe the standard Abel--Jacobi mapping ${\cal
C}^{(g)}\to$Jac$(\cal C)$. Hence, first symmetric polynomial has
the following standard form in terms of theta-functions in the odd
order case
\begin{equation}
\label{along2} \mu_1+\cdots +\mu_{n-1}=c_1
-\partial^2_V\log\theta[\Delta+\eta_{2n}](z-q/2), \qquad
c_1=\sum_{k=1}^{g} \oint_{A_k} \bar{\omega}_k .
\end{equation}
On the other hand, the equation (\ref{2n-kind}) implies that at a
peak point the coordinate $x$ becomes a function of $z$ and
therefore of $t$: $x=p_0(t)$. In the odd order case, this equation
contains a sum of Abelian integrals of the second kind, so-called
{\it Abelian transcendent}. By making use of the following
standard expression for the normalized transcendent (Clebsch and
Gordon [1866])
$$
\sum_{i=1}^{g} \int_{\mu_0}^{\mu_i}\,
\Omega_{\infty}^{(1)} =-\partial_V\log\theta[\Delta+\eta_{2n}]
\left(\sum_{i=1}^{g} \int_{P_0}^{\mu_i}\, \bar{\omega}_k \right) ,
$$
from (\ref{good}) and (\ref{2n-kind}) we find
\begin{equation}
\label{transcendent-2} p_0(t)=\frac 1{\sqrt{-1} L_0} \left(
\partial_V\log\theta[\Delta+\eta_{2n}](z-q/2)-(d,z)+h/2
\right), \qquad
h=\int_{\mathcal{Q}_-}^{\mathcal{Q}_+}\,\Omega_{\infty}^{(1)}.
\end{equation}
Using the trace formula for the solution
$U(p_0(t),t)=\mu_1+\cdots+\mu_{n-1}$ and expression (\ref{along2})
it follows that the equation $x=p_0(t)$ determines a surface
$\mathcal{S}_{0}$ in ${\mathbb C}^n$ along which the solution $U$
has a peak.

Now setting in (\ref{generalized-II}) $P_n\equiv\mathcal{Q}_{-}$
and taking into account  (\ref{norm}), (\ref{norm2}) and
$$
\int_{P_0}^{\mathcal{Q}_-}\,\Omega_{\infty}^{(1)} =-
\int_{P_0}^{\mathcal{Q}_+}\,\Omega_{\infty}^{(1)}, \quad
\int_{P_0}^{\mathcal{Q}_-}\,{\bar\omega} =-
\int_{P_0}^{\mathcal{Q}_+}\,{\bar\omega}
$$
we obtain an
expression for another peak surface $\mathcal{S}_{1}$ determined
by the equation $ \{x=p_1(t)\}$ with
\begin{equation}
\label{trans-next} p_1(t)=\frac 1{\sqrt{-1} L_0} \left(
\partial_V\log\theta[\Delta+\eta_{2n}](z+q/2)-(d,z)-h/2 \right),
\end{equation}
along which
\begin{equation}
\label{along} U(p_1(t),t)=\mu_1+\cdots +\mu_{n-1}=C_1
-\partial^2_V\log\theta[\Delta+\eta_{2n}](z+q/2).
\end{equation}
Under the reality condition, the surfaces $\mathcal{S}_{0}$ and
$\mathcal{S}_{1}$ do not intersect and therefore determine a
connected domain in ${\mathbb C}^n=(x,t)$ where the solution
(\ref{rational}) is applicable. We denote this piece of solution
as $U_1(x,t)$. As follows from (\ref{transcendent-2}) and
(\ref{trans-next}) $\mathcal{S}_{1}$ is obtained from
$\mathcal{S}_{0}$ by changing the phase as follows
\begin{equation}
\label{shift}  Z\to Z+h, \quad z\to z+q \quad {\rm that} \; {\rm
is} \quad x\to x+\frac 1 {\sqrt{-1} L_0}(h-(d,Dt)), \quad t\to t+D^{-1}q,
\quad .
\end{equation}
In addition, according to (\ref{along}) and (\ref{along2}) at any
two points on $\mathcal{S}_{0}$ and $ \mathcal{S}_{1}$ which are
equivalent modulo the shift, $U_1(x,t)$ takes the same values:
\begin{equation}
\label{match} U_1(q_1(t),t) =U_1\left( q_0(t)+\frac 1
{\sqrt{-1} L_0}(h-(d,Dt)),t+D^{-1}q \right) .
\end{equation}
Now let us define function
$U_2(x,t)=U_1(x+(h-(d,Dt))/(\sqrt{-1} L_0),t+D^{-1}q)$, which is also a
local solution to (\ref{dym-sw-eqn}). In view of (\ref{match}),
$U_1$ and $U_2$ take the same values along $\mathcal{S}_1$, which
ensures a correct gluing of two pieces together. By using
iteration with respect to both positive and negative  $N's$, we
construct complete sequence of peak surfaces and obtain formulae
given in part 1) of the theorem.

Similarly, solution (\ref{exponential}) of (\ref{SW-sw-eqn}) is
valid until one of the points $P_1,\dots, P_n$ on $\cal C$
coincides with $\mathcal{Q}_-$ or $\mathcal{Q}_+$, the poles of
$\Omega_0$. Setting $P_n\equiv \mathcal{Q}_+$ in
(\ref{generalized-II}) for the case of an even order curve $\cal
C$, and using (\ref{norm}) and (\ref{norm3}) yields
\begin{align}
\label{good-3} \sum_{i=1}^{n-1}  \int_{P_0}^{\mu_i}\,
\bar{\omega}_k &= z_k- q_k/2 , \qquad k=1,\dots,n-1, \\
\label{3d-kind} \sum_{i=1}^{n-1}  \int_{P_0}^{\mu_i}\,
\left(\Omega_{\infty_\pm} -\sum_{k=1}^{n-1} {\bar
d}_k\bar{\omega}_k\right) &=x - \int_{P_0}^{\mathcal{Q}_+}
\left(\Omega_{\infty_\pm} -\sum_{k=1}^{n-1} {\bar d}_k\bar{\omega}_k\right),
\end{align}
where $P_0=(m_{2n+1},0)$. Inverting  (\ref{good-3}) results in the
following expression for a symmetric polynomial (see e.g., Clebsch
and Gordon [1866])
\begin{equation}
\label{tr} \mu_1+\cdots +\mu_{n-1}
    ={\rm const}-\partial_V \log \frac
{\theta[\Delta](z-{\hat q}-q/2)} {\theta[\Delta](z+{\hat q}-q/2)}.
\end{equation}
After applying the theta-functional formula for the normalized
transcendent of the third kind (Clebsch and Gordon [1866]),
$$
\sum_{i=1}^{g} \int_{P_0}^{\mu_i}\, \Omega_{\infty_\pm} ={\rm
const}- \log \frac {\theta[\Delta](s-{\hat q})}
{\theta[\Delta](s+{\hat q})}, \qquad
s=\sum_{i=1}^{g}\int_{P_0}^{\mu_i}\, \bar{\omega}, \qquad g=n-1,
$$
from (\ref{3d-kind}) and (\ref{good-3}) we obtain
\begin{equation}
\label{q0} x=p_0(t)={\rm const} -\log \frac
{\theta[\Delta](z-{\hat q}+q/2)}
    {\theta[\Delta](z+{\hat q}+q/2)}-(z,{\bar d})+{\bar h}/2 .
\end{equation}
By choosing $P_n\equiv \mathcal{Q}_+$ in (\ref{generalized-II}),
one arrives at the expressions (\ref{tr}) and (\ref{q0}) with
$q/2, {\bar h}/2$ replaced by $-q/2, -{\bar h}/2$. Then, following
similar arguments and applying induction, the piecewise solution
of part 2) is constructed.

We emphasize that although the different pieces $U_N(x,t)$ of the
solution are obtained by iterative shifting the phases $z,Z$ by
the same vector, the pieces $U_N(x,t_0)$ of the solution ($t_0$
being fixed) are all distinct because the shift occurs in both
$x$- and $t$-directions.

\paragraph{Remark.} If we omit the reality condition above, then the
hypersurfaces $\cal S_N$, $\bar{\cal S}_N$ in ${\mathbb C}^n$
intersect, bounding a set of $n$-dimensional domains adjacent to
each other in a rather complicated manner. Then the procedure of
gluing different pieces of the functions $U_N(x,t)$ meromorphic inside
each domain cannot be defined uniquely. As a result, the generic
complex solution $U(x,t)$ branches along the peak surfaces.

\section{Kinematics of peaks.}
\setcounter{equation}{0}
       Now we obtain expressions for the
velocity of $N$-th peak $p_N(t)$ of the piecewise solution of
(\ref{dym-sw-eqn}) with respect to time $t_k$. As was shown above,
the solution has a peak when one of the $\mu$-variables passes
zero implying that $P_n=\mathcal{Q}_-$ or ${P_n=\cal Q}_+$.

\begin{thm}
\label{velocity} Let  $y_1,\dots,y_{n-1}$ denote the
$\mu$-coordinates of the  points $P_1,\dots,P_{n-1}$ at the moment
in time when one of the $\mu$-variables passes zero. The following
system of equations holds:
\begin{equation}
\frac{\partial p_N(t)}{\partial
t_k}=-\Sigma_{k-1}(y_1,\dots,y_{n-1}),
\end{equation}
where $\Sigma_{k-1}$ is the symmetric function defined in
(\ref{higher-t}). In particular, we have
\begin{equation}
\label{qdottrace} \frac{\partial p_N(t)}{\partial t_2}
=y_1+\cdots+y_{n-1}=U(p_N(t),t),
\end{equation}
i.e., the $t_2$-velocity of the peak coincides with its height.
\end{thm}

\noindent{\it Proof.} After applying limit $m_1\to 0$, the
equations (\ref{mux}) and (\ref{higher-t}) for the derivatives of
$\mu_n$ take the form
\begin{eqnarray}
\label{PD-mu} \frac{\partial\mu_n} {\partial x} &=&
\frac{\sqrt{\rho(\mu_n)}}{(\mu_{n}-\mu_{1})\cdots(\mu_{n}-\mu_{n-1})},
\\ \label{PD-mu-k} \frac {\partial\mu_n} {\partial
t_k}&=&\Sigma_{k-1}(\mu_1,\dots,\mu_{n-1})
\frac{\sqrt{\rho(\mu_n)}}{(\mu_{n}-\mu_{1})\cdots(\mu_{n}-\mu_{n-1})}.
\end{eqnarray}
On the other hand, along the peak line $\{x=p_N(t_k)\}$, we have
$$ \frac{\rm d}{\rm d t} \mu_n(p_N(t_k),t_k) \equiv \frac{\partial
\mu_n} {\partial x} \frac{{\rm d}\, p_N(t_k)}{{\rm d}\, t_k}
+\frac{\partial \mu_n}{\partial t_k}=0, $$ which, in view of
(\ref{PD-mu}) and (\ref{PD-mu-k}) and after setting $\mu_n\equiv
0$, yields $$ \frac{\sqrt{\rho(0)}}{\mu_{1}\cdots \mu_{n-1} }
\left ( \frac{\partial \mu_n}{\partial
t_k}+\Sigma_{k-1}(y_1,\dots,y_{n-1}) \right)=0. $$ Since
$\rho(0)\ne 0$ and $\mu_1=y_1,\dots,\mu_{n-1}=y_{n-1}$ are finite,
     the latter relation gives (\ref{velocity}).

\paragraph{Remark.}  The relations (\ref{velocity}) can be also found by
using direct differentiation of the expression for the $N$-th peak
surface (\ref{odd-peak-surface}) with respect to  $t_k$. Namely,
putting without loss of generality $N=0$, and taking into account
$\partial_V=\partial_{t_2}$, we write
\begin{equation}
\frac{\partial p_0(t)}{\partial t_k}=\partial_{t_k} \partial_{t_2}
\log\theta[\Delta+\eta_{2n}] \left(\sum_{i=1}^{n-1}
\int_{P_0}^{\mu_i}\, \bar{\omega} \right) .
\end{equation}
According to Mumford [1983], in case of odd order hyperelliptic
curves, this gives a theta-functional expression for the
coefficient in front of $\lambda^{n-k}$ in the polynomial
$(\lambda-\mu_1)\cdots (\lambda-\mu_{g})$ which coincides with
$\Sigma_{k-1}(y_1,\dots,y_{n-1})$.

\section{ The dynamics of peaks and weak
solutions.} \setcounter{equation}{0}

Expression~(\ref{qdottrace}) states that for equations in the
hierarchies of (HD) or (SW), every peak in the solution profile
move with velocity determined by the local value of the solution.
In this section, we derive this property without recourse to tools
related to the complete integrability of the evolution equation.
Thus, this property of peak motion can hold in general for
equations that admit piecewise-smooth weak solutions, with jumps
in the first spatial derivative at isolated points in the
solution's support. In this case, the derivative discontinuity can
be viewed as a `shock' in the appropriate weak form of the
evolution equation.

    We will take
the weak form of the equation (HD) or (SW) to be
\begin{equation} \label{chsw4}
{\displaystyle \int\limits_{\Omega}  \nabla\phi(x,t)  \cdot {\bf
V}(x,t) \, \,  dx \; dt = 0}  \, ,
\end{equation}
where the equality is satisfied for all test functions $\phi
(x,t)$ is $C^\infty$ with compact support in a domain $\Omega$ in
the $(x,t)$ plane. Here $\nabla \phi =(\phi_t, \phi_x) $, the dot
denotes $\mathbb{R}^2$ inner product, and the vector function
${\bf V}(x,t)=(V_1, V_2)$ is defined by
\begin{align}\label{vhddef}
V_1 & = U_x \nonumber
\\
V_2 & = \partial_x \left[ {\frac{1}{2}}U^2 -{\frac{1}{4}} \int_{-
\infty}^{\infty}  \;
    |x-y| \; (U_{y}^2 - 2\kappa U)\; dy \right] \, ,
\end{align}
for equation (HD) and
\begin{align}\label{vswdef}
V_1 & = U_x \nonumber
\\
V_2 & =  \partial_x \left[ {\frac{1}{2}}U^2 + {\frac{1}{4}}\int_{-
\infty}^{\infty}  \; e^{-|x-y|} \; \left(2 U^2 + U_{y}^2 -2 \kappa
U\right)\; dy \right] \, ,
\end{align}
for equation (SW), respectively. We will look for jump conditions
satisfied by the solutions of equation~(\ref{chsw4}). If the jump
discontinuities are isolated, by adjusting the support of the test
functions $\phi(x,t)$ we only need to consider the case of a
single discontinuity.

Let us suppose that the function $U(x,t)$ is infinitely
differentiable almost everywhere in $\Omega$, except along the
curve $x=q(t)$ where the first derivative $U_x$ has a
discontinuity. If we partition the domain $\Omega$ into
$\Omega=\Omega_1\cup \Omega_2$ by cutting along the portion of the
discontinuity curve $x-q(t)=0$ in $\Omega$,   the divergence
theorem and the choice of test functions $\phi(x,t)$ vanishing on
the boundary $\partial \Omega$ allow us  to write equation
(\ref{chsw4}) as
\begin{equation} \label{chsw41}
0= \int\limits_{\Omega} \;  dx \; dt \; \nabla\phi \, \cdot {\bf
V} = \int\limits_{\Omega_1} \;  dx \; dt \; \phi \, \nabla \cdot
{\bf V} + \int\limits_{\Omega_2} \;  dx \; dt \; \phi \, \nabla
\cdot {\bf V} + \hspace{-0.2cm}\int \limits_{\partial \Omega_1
\cap \partial \Omega_2 } \hspace{-0.3cm}d l \, \phi \,  {\bf
n}\cdot [{\bf V}]_-^+ \, .
\end{equation}
Here the unit vector $\bf n$ is directed along the normal
$[-\dot{q}, 1]$ to the discontinuity  curve ${\partial \Omega_1
\cap \partial \Omega_2 }$ in $\Omega$, and $[{\bf V}]_-^+$ denotes
the jump of the vector $\bf V$ across this  curve, $$ [{\bf
V}]_-^+ \equiv \lim_{x \to q(t)^+}{\bf V}(x,t) -\lim_{x \to
q(t)^-}{\bf V}(x,t) \, . $$

By the arbitrariness of $\phi(x,t)$, each integrand term on the
right-hand side of~(\ref{chsw41}) has to vanish separately. Thus,
from the first two terms,
\begin{equation} \label{divvfrm}
\nabla \cdot {\bf V} = 0 \, ,
    \qquad \hbox{or}  \qquad
\frac{\partial  V_{1}}{\partial t} + \frac{\partial
V_{2}}{\partial x} = 0 \, ,
\end{equation}
in $\Omega_1$ or $\Omega_2$, where $U(x.t)$ is smooth. This
smoothness and zero divergence condition, by the definition
(\ref{vhddef}) or (\ref{vswdef}) for (HD) or (SW) respectively,
imply that $U(x,t)$ is a solution of these equations in $\Omega_1$
or $\Omega_2$. For instance, (\ref{divvfrm}) becomes $$
U_{xt}+\partial_{xx}\left[ {\frac{1}{2}}U^2 -{\frac{1}{4}} \int_{-
\infty}^{\infty}  \;
    |x-y| \; (U_{y}^2 - 2\kappa U)\; dy \right] =0
$$ which is the integrated form of the Harry-Dym equation (HD).

The last (jump)  condition in~(\ref{chsw41}), ${\bf n}\cdot  [{\bf
V}]_-^+$ =0 along $\partial \Omega$, implies
\begin{equation}\label{jump}
\dot{q}[V_1]_-^+=[V_2]_-^+ \, .
\end{equation}
The left-hand side of this expression is simply
\begin{equation}\label{v1jump}
\dot{q}[V_1]_-^+= \dot{q}[U_x]_-^+ \, .
\end{equation}
As to the right-hand side, the second (integral) term in the
definitions (\ref{vhddef}) or (\ref{vswdef}) of $V_2(x,t)$ is a
continuous function of $x$, as the integral wipes out the
discontinuity $\operatorname{sgn}(x-y)$ as well as additional ones
that $U^2_y$ might have.  Hence the integral terms do not
contribute  to the right hand side of (\ref{jump}). The jump of
$V_2(x,t)$ across the discontinuity curve $x=q(t)$ then reduces to
\begin{equation}\label{v2jump}
[V_2]_-^+={\frac{1}{2}} \left[\left(U^2)_x\right)\right]_-^+
=U(q,t)\,[U_x]_-^+ \, .
\end{equation}
If $$ [U_x]_-^+ \equiv U_x (q^+ , t) - U_x (q^- , t) \neq 0 $$
equations (\ref{v1jump}) and (\ref{v2jump}) yield
\begin{equation}\label{sspeed}
\dot q = U(q,t) \, ,
\end{equation}
i.e., the location of the discontinuity (shock) in the $U_x$ moves
at the local speed $U(q,t)$. We have then proved the following
\begin{thm}
\label{velocity.pde} Let $U(x,t)$ be a solution of equation
(\ref{chsw4}), with the vector ${\bf V}(x,t)$ defined in terms of
$U(x,t)$ by the nonlinear, nonlocal operators (\ref{vhddef}) and
(\ref{vswdef}) respectively for equations (HD) and (SW). Let
$U(x,t)$ be a smooth function of $(x,t)$ in the domain $\Omega
\subseteq \mathbb{R}^2$, except along the curve $x=q(t)$, where
$U$ is continuous while the first derivative $U_x$ has a jump
discontinuity (peak) $U(q^+,t)\neq U(q^-,t)$. Then $U(x,t)$ is a
solution of equations (HD) and (SW) in each domain $\Omega_1$ and
$\Omega_2$ in which the curve $x=q(t)$ partitions $\Omega$, and
the location of the peak $q(t)$ moves with velocity equal to its
height, $\dot{q}= U(q,t)$.
\end{thm}

\paragraph{Conclusions.}
In this paper,  profiles of the weak finite-gap piecewise-smooth solutions
of the integrable nonlinear equations of shallow water and Dym type
are linked to billiard dynamical systems and geodesic flows with reflections
described in terms of finite dimensional Hamiltonian systems on Riemann
surfaces. After reducing the solution of these systems to
that of a nonstandard Jacobi inversion problem, solutions
are found by introducing new parametrizations. The extension of the
algebraic-geometric method for nonlinear integrable PDE's given in this paper
leads to description of piecewise-smooth weak solutions of a class of
$N$-component systems of nonlinear evolution equations and its
associated energy dependent Schr\"{o}dinger operators.

\paragraph{Acknowledgments.}
Mark Alber and Roberto Camassa would like to thank Francesco
Calogero and Al Osborne for helpful discussions.  The authors
would like to thank R. Beals, D. Sattinger and J. Szmigielski for
pointing out their recent work and for making it available.

\subsection*{References}

\begin{description}

\item Abenda, S., Fedorov, Yu. [2000] On the weak
Kowalevski--Painlev\'e property for hyperelliptically separable
systems. {\it Acta Appl. Math.} {\bf 60} (2)  137--178.

\item  Ablowitz, M.J.,  Segur, H. [1981]
{\it Solitons and the Inverse Scattering Transform}. Philadelphia:
SIAM.

\item Alber, S.J. [1979]  Investigation of equations of Korteweg
de Vries type by the method of recurrence relations. (Russian)
{\it J. London Math. Soc.} (2) {\bf 19}, no. 3, 467--480

\item Alber, M.S.,  Alber, S.J. [1985]  Hamiltonian formalism for
finite-zone solutions of integrable equations. {\it C. R. Acad.
Sci. Paris Ser.I Math.}  {\bf 301}, 777--781.

\item Alber, M.S., Camassa, R.,  Holm, D.D. and  Marsden, J.E. [1994]
The geometry of peaked solitons and billiard solutions of a class
of integrable PDE's. {\it Lett. Math. Phys.} {\bf 32}, 137--151.

\item Alber, M.S., Camassa, R.,  Holm, D.D. and  Marsden, J.E. [1995]
On the link between umbilic geodesics and soliton solutions of
nonlinear PDE's. {\it Proc. Roy. Soc.} {\bf  450}, 677--692.

\item Alber, M.S., Camassa, R., Fedorov, F., Holm, D.D. and  Marsden, J.E.
[1999] On billiard solutions of nonlinear PDE's, {\it Phys. Lett.
A} {\bf 264}, 171--178.

\item Alber, M.S.  and Miller, C. [2001] On peakon solutions of the
shallow water equation, {\it Appl.Math. Lett.}  {\bf 14}, 93--98.

\item Alber, M.S. and Fedorov, Yu.N. [2000]  Wave solutions of
evolution equations and Hamiltonian flows on nonlinear
subvarieties of generalized Jacobians, {\it J. Phys.A: Math. Gen.}
{\bf 33}, 8409--8425.

\item Alber, M.S. and Fedorov, Yu.N. [2001]  Algebraic geometric solutions
for nonlinear evolution equations and flows on the nonlinear
subvarieties of Jacobians,  {\it Inverse Problems} (to appear)

\item Alber, M.S., Luther, G.G., and Marsden, J.E. [1997]
Complex billiard Hamiltonian systems and nonlinear waves. In:
Fokas, Y.H.,  Gelfand, I. M. (eds.) {\it Algebraic Aspects of
Integrable Systems.} Boston: Birkh\H{a}user.

\item Antonowicz, M. and Fordy, A. P. Factorization of energy
dependent Schr\"{o}dinger operators: Miura maps and modified systems. Comm.
Math. Phys. 124 (1989), no. 3, 465--486.

\item  Beals, W., Sattinger, D., and Szmigielski, J. [1998]
Acoustic scattering and the extended Korteweg de Vries hierarchy.
{\it Adv. Math.} {\bf 140}, 190-206.

\item Beals, R., D.H. Sattinger, and J. Szmigielski [1999],
Multi-peakons and a theorem of Stietjes, {\it Inverse Problems}
{\bf 15} L1--L4.

\item Beals, R., Sattinger,  D.H. and Szmigielski, J. [2000],
Multipeakons and the classical moment, {\it Advances in
Mathematics} {\bf 154}, no. 2, 229--257.

\item Belokolos, E.D., Bobenko, A.I., Enol'sii, V.Z.,  Its, A.R., and
Matveev, V.B. [1994] {\it Algebro-Geometric Approach to Nonlinear
Integrable Equations}. Springer-Verlag series in Nonlinear
Dynamics. New York: Springer-Verlag.

\item Bobenko, A. I. and Suris, Y. B. [1999] Discrete Lagrangian reduction,
discrete Euler-Poincar\'e equations, and semidirect products. {\it
Lett. Math. Phys.} {\bf 49} , no. 1, 79--93.

\item Bulla, W., Gesztesy, F., Holden, H. and Teschl, G.
[1998] {\it Algebro-geometric quasi-periodic finite-gap solutions of the Toda
and Kac-van Moerbeke hierarchies}. Mem. Amer. Math. Soc. {\bf 135}.

\item Camassa, R. [2000] Characteristic variables for a completely
integrable shallow water equation. In: Boiti, M. et al. (eds.)
{\it Nonlinearity, Integrability  and All That: Twenty Years After
NEEDS '79.} Singapore: World Scientific.

\item Camassa, R., Holm, D.D. [1993]  An integrable
shallow water equation with peaked solitons. {\it  Phys. Rev.
Lett.} {\bf 71}, 1661--1664.

\item Camassa, R., Holm, D.D. and Hyman, J.M. [1994]
A new integrable shallow water equation. {\it Adv. Appl. Mech.},
{\bf 31}, 1--33.

\item Calogero, F. [1995] An integrable Hamiltonian system.
{\it Phys. Lett. A.} {\bf 201}, 306--310.

\item Calogero, F. and Francoise, J.-P. [1996] Solvable quantum
version of an integrable Hamiltonian system. {\it J. Math. Phys.}
{\bf 37} (6),  2863--2871.

\item Cewen, C. [1990] Stationary Harry-Dym's equation and its
relation with geodesics on ellipsoid. {\it Acta Math. Sinica} {\bf
6}, 35--41.

\item Clebsch, A. and Gordon, P. [1866] {\it Theorie der abelschen
Funktionen.} Teubner, Leipzig.

\item Dmitrieva, L.A. [1993a] Finite-gap solutions of the Harry Dym
equation, {\it Phys.Lett. A} {\bf 182}, 65--70.

\item Dmitrieva, L.A. [1993b] The higher-times approach to multisoliton
solutions of the Harry Dym equation,  {\it J. Phys. A} {\bf 26}
6005--6020.

\item Drach, M. [1919] Sur l'integration par quadratures de l'equation
$d^2y/dx^2=[\phi(x)+h]y$. {\it Comptes rendus} {\bf 168}, 337--340

\item Dubrovin,  B.A. [1975] Periodic problems for the Korteweg-de Vries
equation in the class of finite-band potentials. {\it Funct. Anal.
Appl.} {\bf 9} 215--223.

\item Dubrovin,  B.A. [1981] Theta-functions and nonlinear equations.
{\it Russ. Math. Surv.} {\bf 36} (2), 11--92.

\item  Ercolani, N. [1987] Generalized theta functions and
homoclinic varieties. In: Ehrenpreis, L., Gunning, R.C. (eds.)
{\it Theta functions}. Proceedings, Bowdoin. 87--100. Providence,
R.I.: American Mathematical Society.

\item Fedorov, Yu. [1999] Classical integrable systems related to
generalized Jacobians. {\it Acta Appl. Math.} 55, 3, 151-201.

\item Fedorov, Yu. [2001] Ellipsoidal billiard with the quadratic
potential, {\it Funct. Anal. Appl.} (Russian) (to appear).

\item Gavrilov, L. [1999] Generalized Jacobians of spectral curves and
completely integrable systems. {\it Math. Z.}, {\bf 230}, 487--508.

\item Gesztesy, F. [1995] New trace formulas for SchrÃdinger operators.
Evolution equations (Baton Rouge, LA, 1992), 201--221, {\it Lecture Notes in
Pure and Appl. Math.}, {\bf 168}, Dekker, New York.

\item Gesztesy, F. and Holden, H. [1994] Trace formulas and conservation laws
for nonlinear evolution equations. {\it Rev. Math. Phys.} {\bf  6},
51--95 and 673.

\item Gesztesy, F. and Holden, H. [2001] Algebraic-geometric
solutions of the Camassa-Holm equation,  (preprint).

\item Gesztesy, F. and Holden, H. [2001] Dubrovin equations and integrable
systems on hyperelliptic curves. {\it Math. Scand.} (to appear).

\item Gesztesy, F., Ratnaseelan, R., and Teschl, G. [1996] The KdV
hierarchy and associated trace formulas. Recent developments in operator
theory and its applications (Winnipeg, MB, 1994), 125--163, Oper. Theory
Adv. Appl., 87, Birkhauser, Basel, 1996.

\item Hunter, J.K. and  Zheng, Y.X. [1994] On a completely
integrable nonlinear hyperbolic variational equation. {\it Physica
D} {\bf 79}, 361--386.

\item  Jacobi, C.G.J. [1884a] {\it Vorlesungen uber Dynamik,
Gesamelte Werke}. Berlin: Supplementband.

\item Jacobi, C.G.J. [1884b]
Solution nouvelle d'un probleme fondamental de geodesie. Berlin:
{\it Gesamelte Werke}  {\bf Bd. 2}.

\item Jaulent, M. [1972] On an inverse scattering problem with an
energy dependent potential. {\it Ann. Inst. H. Poincare A} {\bf
17}, 363--372.

\item Jaulent, M. and Jean, C. [1976] The inverse problem for the
one-dimensional Schr\"{o}dinger operator with an energy dependent
potential. {\it Ann. Inst. H. Poincare A. I, II} {\bf 25},
105--118, 119--137.

Kane, C., E. A. Repetto, E.A.,  Ortiz, M., and Marsden, J.E.  [1999],
Finite element analysis of nonsmooth contact, {\it Comput. Methods
      Appl. Mech. and Engrg.}, {\bf 180}, 1-26.

\item Klingerberg, W. [1982] {\it  Riemannian Geometry}.
New York: de Gruyter.

\item Kn\"{o}rrer, H. [1982]
Geodesics on quadrics and mechanical problem of C. Neumann. {\it
J. Reine Angew. Math.} {\bf 334}, 69--78.

\item Kozlov, V.V. and Treschev,  D. V. [1991] {\it Billiards, a
Genetic Introduction to Systems with Impacts}. AMS Translations of
Math. Monographs {\bf 89}. New York.

\item Kruskal, M.D. [1975] Nonlinear wave equations. In Moser, J.
(eds.) {\it Dynamical Systems, Theory and Applications}. Lecture
Notes in Physics {\bf 38}, New York: Springer.

\item  Markushevich, A. I. [1977] {\it Introduction to the Theory of
Abelian Functions}. English version: Translations of Mathematical
Monographs, 96. American Mathematical Society, Providence, RI,
1992.

\item Marsden, J. E.,  Patrick, G.W., and  Shkoller, S.  [1998]
Mulltisymplectic geometry, variational integrators and nonlinear
PDEs, {\it Comm. Math. Phys.}, {\bf 199}, 351--395.

\item Marsden, J. E.,  Pekarsky, S., and  Shkoller, S. [1999] Discrete
Euler-Poincare and Lie-Poisson equations, {\it Nonlinearity} {\bf
12}, 1647--1662.

\item Marsden, J. E. and Ratiu, T.S. [1999] {\it Introduction to
Mechanics and Symmetry}, Texts in Applied Mathematics, {\bf 17},
Springer-Verlag.

\item McKean, H.P. and Constantin, A. [1999] A shallow water equation on
the circle, {\it Comm. Pure Appl. Math.} {\bf Vol LII}, 949--982.

\item Moser, J. and  Veselov, A.P. [1991]
Discrete versions of some classical integrable systems and
factorization of matrix polynomials.  {\it Comm. Math. Phys.\/}
{\bf 139}, 217--243.

\item Mumford, D. [1983]
{\it Tata Lectures on Theta I and II.} Boston: Birkhauser-Verlag.

\item Novikov D.P. [1999] Algebraic-geometrical solutions of the
Harry--Dym equations. {\it Sibirskii Matematicheskii Zhurnal},
{\bf 40} 159--163, (Russian) English transl. in: {\it Siberian
Math. Journal}, {\bf 40}, 136--140.

\item Previato, E. [1995] Hyperelliptic quasi-periodic and soliton
solutions of the nonlinear Schr\"{o}dinger equation. {\it Duke
Math. Journal} {\bf 52}, 329--377.

\item Rauch-Wojciechowski, S. [1995] Mechanical systems related to
the Schr\"{o}dinger spectral problem. {\it Chaos, Solitons $\&$
Fractals} {\bf 5}, 2235--2259.

\item Serre, J.-P. [1959] {\it Groupes alg\'ebriques et corps de
classes} Hermann, Paris.

\item Vanhaecke, P. [1995] Stratification of hyperelliptic Jacobians
and the Sato Grassmannian. {\it Acta Appl. Math.} {\bf 40},
143--172.

\item Veselov, A.P. [1988]
Integrable discrete-time systems and difference operators. {\it
Funct. An. and Appl.\/} {\bf 22}, 83--94.

\item Verhulst, F.  [1996]
{\it Nonlinear Differential Equations and Dynamical Systems}.
Second Edition. Springer-Verlag.

\item Wadati, M.,  Ichikawa, Y.H., and Shimizu, T. [1980] Cusp soliton
of a new integrable nonlinear evolution equation. {\it Prog.
Theor. Phys.} {\bf 64}, 1959-1967.

\item Weierstrass, K. [1884]
\"Uber die geod\"atischen Linien auf dreiachsigen Ellipsoid, {\it
Math. Werke I}, 257--266.

\item Wendlandt, J.M. and Marsden, J.E. [1997]  Mechanical
integrators derived from a discrete variational principle,  {\it
Physica D\/} {\bf  106}, 223--246.

\item Whittaker, E.T. [1937] {\it A Treatise on
the Analytical Dynamics of Particles and Rigid Bodies\/},
Cambridge University Press, 1904; 4th Ed., 1937  (reprinted by
Dover 1944, and Cambridge University 1988.)

\item Young, L.C. [1969] {\it Lectures on the Calculus of Variations and
Optimal Control Theory}, Saunders, Corrected printing, Chelsea,
1980.

\end{description}

\end{document}